has been experimentally exploredBy now,
\documentclass[twocolumn,showpacs,preprintnumbers,amsmath,amssymb,nofootinbib]{revtex4}

\usepackage{graphicx}% Include figure files
\usepackage{dcolumn}% Align table columns on decimal point
\usepackage{bm}% bold math
\usepackage{amssymb}
\usepackage{amsmath}
\usepackage{amsthm}
\begin{document}

%\preprint{APS/123-QED}

\title{Mean volume reflection angle% for silicon crystal in (110) and (111) orientations
}

\author{M. V. Bondarenco}
 \email{bon@kipt.kharkov.ua}
%\email{bon@kipt.kharkov.ua}
%\author{N.F.~Shul'ga}
 \affiliation{%
NSC Kharkov Institute of Physics and Technology, 1 Academic St.,
61108 Kharkov, Ukraine }
 \affiliation{%
V.N. Karazine Kharkov National University, 4 Svobody Sq., 61077
Kharkov, Ukraine }

\date{\today}% It is always \today, today,

\begin{abstract}

The mean volume reflection angle of a high-energy charged particle passing through a bent crystal is expressed as an integral involving the effective interplanar potential over a single crystal period. Implications for positively and negatively charged particles, and silicon crystal orientations (110) and (111) are discussed. 
A generic next-to-leading-order expansion in the ratio $E/R$ of the particle energy $E$ to the crystal bending radius $R$ is given. 
For positively charged particles, the dependence of the mean volume reflection angle on $E/R$ proves to be approximately linear, whereas for negatively charged particles the linear behaviour is modified by an $E/R$-dependent
%sizeable 
logarithmic factor. 
Up-to-date experimental data are confronted with predictions based on commonly used atomic potentials.

%For positively charged particles, the volume reflection angle is found to have a discernible dependence on the crystal temperature.

%Dimuon registration efficiency is evaluated in analytic form.

\end{abstract}

                             


%\end{keyword}

\maketitle

\section{Introduction}\label{sec:Intro}

Volume reflection (VR) is deflection of high-energy charged particles by a planarly oriented bent crystal. It may be regarded as an effect complementary to channeling in bent crystals, in which particles deflect to the side opposite to that of the crystal bending, whereas channeled particles deflect towards the crystal bending. 
It is considered to be applicable for beam steering at multi-GeV accelerators (e.g., in a multiple-VR mode), having the merit of high acceptance (see \cite{Scand-Tar-PhysRep} and refs. therein). 
By now, it has been experimentally explored for various bent crystal orientations, incident charged particle types and beam energies \cite{Ivanov1GeV,Ivanov70GeV,Scandale,Scandale-PRST-2008,ScandalePLB111,Scandale-neg,Hasan,Rossi-Scandale-2015,Wistisen-Wienands,Bandiera-VR-rad}. 

VR effect owes to the asymmetry of the effective continuous potential of atomic planes caused by the crystal bending, but it does not vanish when this asymmetry becomes small. 
On the contrary, in that case it becomes maximal. 
Its theoretical treatment is facilitated by the fact that if, as is normally the case, the bending is sufficiently uniform, then the particle angular momentum (or transverse energy) in the effective potential with a centrifugal component is conserved \cite{Tar-Vor}. 
That makes the problem integrable and in principle analytically tractable \cite{Maisheev,Kovalev,Bond-VR-PRA,Bond-PLA,Bond-PRST,Shulga-Truten-Boyko-Esaulov-VR,Bellucci-Chesnokov-Maisheev-Yazynin}. 
However, since VR angle is accumulated over many interplanar intervals, in each of which the centrifugal potential component is different, yet the interplanar potential itself is generally given by a sophisticated function, the VR angle dependencies on the particle energy and charge sign, crystal material, orientation and bending radius are obscured, ultimately demanding numerical evaluation. 

To get a grasp of various parameter dependencies in the VR problem, in \cite{Bond-VR-PRA} it was solved for the simplest example -- parabolic model for the interplanar potential. 
It was demonstrated that the complicated dependencies of the VR angle on $R$ and $E$ greatly simplify if an expansion in the small parameter $R_c/R$ [with $R_c(E)$ being the critical radius] is carried out. 
With its aid, explicit expressions for the VR angle of positively and negatively charged particles were obtained. 
But the model treatment was only adequate for silicon crystal in orientation (110), and did not cover other practically important cases, such as silicon in orientation (111), and germanium crystals. 
%, it has always been tempting to describe this process analytically. In \cite{Bond-VR-PRA,Bond-PRST} explicit formulae were obtained for the VR angular distribution as a function of the particle energy and the crystal curvature radius, which were compared favorably with experimental data. The approach of \cite{Bond-VR-PRA}, however, described the interplanar potential only within a simple parabolic model, and did not treat the case of a practically important orientation (111), so it could not be regarded as quite general. 

The aim of the present paper is to generalize the analytic theory of VR, and develop theoretical tools valid for arbitrary interplanar potential and any crystal orientation. 
To this end, it is expedient to interchange the procedures of radial integration and averaging over transverse energy, which leads to an expression for the VR angle as an integral over a single crystal period of a square root of the effective potential. 
With its aid, a generic expansion in $R_c/R$ can be derived, in which the entire dependence on the straight-crystal potential enters to the coefficient functions. 
Cases of $R\sim R_c$ and $R\to R_c$ can be studied, as well. 
%reflecting the potential shape more accurately, and including the case of (111) to consideration. 
%A new important result is that the VR angle may be expressed as an integral over a single interplanar interval for arbitrary shape of the continuous potential, which greatly facilitates the calculations, and allows one to avoid model assumptions. 
%It will be complemented by calculations for realistic interplanar potentials, comparison with the available experimental data, and by some illuminating model results.  

For simplicity, herein we restrict ourselves to the thick-crystal limit, when contributions from the crystal boundaries are vanishing, and presume the so-called statistical equilibrium, i.e., a uniform distribution in the fast particle transverse energy \cite{Maisheev}. 
Furthermore, we will focus on issues most important for practice, i.e., 
calculation of the mean VR angle, which is often directly extracted from the experimental data, being almost unaffected by incoherent multiple scattering, and application to silicon crystals, usually used in experiments. 
%The analysis of VR angular distributions, and application to germanium crystals will be presented elsewhere. 

In order to assess the accuracy of the obtained exact and approximate representations, we confront their predictions based on popular realistic potentials with the up-to-date world data for silicon bent crystals in orientations (110) and (111), and for both particle charge signs. For the benefit of the reader, we also include a summary of results for a few model potentials, qualitatively illustrating dependencies on their typical parameters.

%Benefiting from the facility of computation of mean reflection angles via the given representations, it is instructive next to investigate its sensitivity to the shape of the interplanar potential. 
%In \cite{Maisheev} it was found that predictions for this observable based on the Moli\`{e}re potential frequently used for other particle-crystal interaction problems were at variance with experimental data for VR of positively charged particles. In fact, a shallower potential was favoured. 
%To explore this issue in more detail, we compare

%first derive closed-form expressions for the mean VR angle in  (parabolic and joint-parabolic) potentials and  on the VR angle. 
%Thereafter, we analyse applications of the Moli\`{e}re potential and its modifications to (110) and (111) orientations of the bent crystal, and ... whether they are reconcilable with experimental data. 

\section{General considerations}\label{sec:gen-framework}

When a fast charged particle traverses a bent crystal, the angle $\theta$ between its velocity and the family of weakly bent and long atomic planes gradually changes. 
At its relatively small values, relevant for the VR effect, the particle motion is governed by relativistic classical mechanics in the continuous potential averaged along the planes, similarly to the case of channeling. 
The period of the force acting on the particle is then tapering to both sides away from the VR region, and in remote regions, in which  $|\theta|\gg\theta_c$, where $\theta_c=\sqrt{2V_0/E}$ is the critical channeling angle for a straight crystal with well depth $V_0$ , its net deflecting action tends to zero. 
As for the incoherent multiple scattering, its strength per unit particle path length remains nearly constant everywhere in the crystal (an $R$-dependent correction to it was evaluated in \cite{Bond-PLA,Bond-PRST}), so, in a thick crystal it eventually becomes formidable. 
But within the intrinsic VR region, it still remains minor compared to the action of the continuous potential. 
Besides that, the incoherent multiple scattering is symmetric wrt the initial particle motion direction. Therefore, in the first approximation it should not contribute tot the mean VR angle, and may be neglected. 

As was pointed out already in paper \cite{Tar-Vor}, where VR was predicted, basic notions facilitating the theoretical description of VR are the same as for channeling in a bent crystal  \cite{Kaplin-Vorobiev}:
\begin{enumerate}
\item
The continuous potential of a uniformly bent crystal in a planar orientation is axially symmetric. 
Thus, it conserves the angular momentum projection on the symmetry axis, or, equivalently, the transverse energy including the centrifugal potential. 
This second integral of the 2-dimensional motion (excluding the irrelevant uniform motion along the crystal bending axis and the planes) makes the problem completely integrable.
\item
For a small crystal bending angle (not damaging the ideal crystal lattice), the centrifugal potential may be linearized within the entire crystal volume.
\end{enumerate}
The particle trajectory in polar coordinates $\{r,\varphi\}$ (radius and angle wrt the bent crystal symmetry axis, located far outside of the crystal) then expresses in an explicit integral form \cite{Tar-Vor}:
\begin{equation}\label{angle-wrt-centre}
    \varphi\left(r,r_0,E_{\perp}\right)
= \frac1R\int_{r}^{r_0}\frac{dr}{\sqrt{2\frac{E_{\perp}-V(r)}{E}+2\frac{r}{R}}}.
\end{equation}
Here $R$ designates the crystal bending radius, $E$ -- the energy of the fast particle (assumed to be ultrarelativistic\footnote{\label{foot:Ev2}In a generic case, $E$ in Eq. (\ref{angle-wrt-centre}) must be replaced by $Ev^2/c^2$, where $v$ is the particle velocity and $c$ the speed of light, but insofar as VR is usually applied to ultrarelativistic particles, for brevity we neglect the difference between $v$ and $c$. For non-ultrarelativistic particles, the corresponding substitution is due in the final equations.}), $E_{\perp}$ -- the particle transverse energy (defined here to include all the $r$-independent terms), $V(r)$ -- the periodic (with period $d$) continuous potential of the corresponding family of atomic planes, which in a uniformly bent crystal depends only on the radial coordinate $r$ normal to the planes, and $r_0$ is the $r$ value at the particle entrance to the crystal.

Eq. (\ref{angle-wrt-centre}) serves as a starting point for all the subsequent calculations. Pure VR experiments, though, measure not the entire particle trajectory inside the crystal, but only the final deflection angle. 
It can be deduced from (\ref{angle-wrt-centre}) as\footnote{Compared to \cite{Maisheev,Bond-VR-PRA}, we define here angle $\chi$ of deflection wrt the initial particle direction of motion with the opposite sign, in order to make it positive. Notations $\theta$ and $\theta_c$ are reserved for angles of particle motion wrt atomic planes.} \cite{Maisheev}
\begin{subequations}\label{eq2ab}
\begin{eqnarray}\label{Maisheev-repr}
    \chi\left(r_0,E_{\perp}\right)
= \frac2R 
\int_{r_c}^{r_0} dr \Bigg\{ \frac{1}{\sqrt{2\frac{E_{\perp}-V(r_c)}{E}+2\frac{r}{R}}}
 \qquad \nonumber\\
-  \frac{1}{\sqrt{2\frac{E_{\perp}-V(r)}{E}+2\frac{r}{R}}} \Bigg\},
\end{eqnarray}
where $r_c$ is the radial reflection point, in which $\frac{E_{\perp}-V(r_c)}{E}+\frac{r_c}{R}=0$, and factor of 2 in Eq. (\ref{Maisheev-repr}) accounts for contributions before and after the radial reflection, assuming the crystal to be oriented symmetrically wrt the beam direction. 
If the first term in the braces in Eq. (\ref{Maisheev-repr}) is integrated explicitly \cite{Bond-VR-PRA,Shulga-Truten-Boyko-Esaulov-VR}, viz.,
\begin{equation}\label{BondShulgaTrutenBoyko}
    \chi
= 2\sqrt{2\frac{E_{\perp}-V(r_0)}{E}+2\frac{r_0}{R}}
-
\frac2R 
\int_{r_c}^{r_0} \!\frac{dr}{\sqrt{2\frac{E_{\perp}-V(r)}{E}+2\frac{r}{R}}} ,
\end{equation}
\end{subequations}
it is evident that the first term 
\[
2\sqrt{2\frac{E_{\perp}-V(r_0)}{E}+2\frac{r_0}{R}}=2\frac{dr_0}{Rd\varphi}
\]
represents the doubled angle between the particle velocity $d\vec{r}/dt$ (with time differential $dt=Rd\varphi$) and the bent aligned atomic planes at the exit from the crystal, whereas the second term -- the polar angle $\varphi$ subtended by the particle trajectory in the crystal, i.e., in effect, the total crystal bending angle. Their difference, naturally, equals the deflection angle in the inertial laboratory frame.

Next, it needs to be taken into account that in practice, the initial state represents not a single particle with a perfectly known impact parameter and velocity, but a beam. 
Particles from the incident beam enter the crystal with random (at an atomic scale) impact parameters and with slightly randomized angles. Accordingly, $E_{\perp}$ is a randomly distributed variable, too. 
One has therefore to derive from (\ref{eq2ab}) the angular distribution of the scattering probability.
Its calculation is alleviated \cite{Maisheev} by relying on the so-called ``statistical equilibrium'' property, i.e., a uniform distribution in $E_{\perp}$ within the relatively small (compared with the initial $E_{\perp}$ uncertainty)\footnote{For a more detailed analysis of conditions of its validity, see \cite{Bond-VR-PRA}.} period $\Delta E_{\perp}={Ed}/{R}$:
\begin{equation}\label{dwdthetaVR=dEperpdthetaVR}
\frac{dw}{d\chi}=\frac{1}{\Delta E_{\perp}}\frac{dE_{\perp}}{d\chi}.
\end{equation}
Therewith, the mean VR angle, on which we will focus in the present paper, is given by a simple expression
\begin{eqnarray}\label{const+DeltaEperp}
\overline{\chi}&=&\int d\chi \frac{dw}{d\chi}\chi\nonumber\\
&=&\frac1{\Delta E_{\perp}}\int_{\text{const}}^{\text{const}+\Delta E_{\perp}} dE_{\perp} \chi(E_{\perp})
\end{eqnarray}
directly in terms of $\chi(E_{\perp})$ rather than its inverse function entering Eq. (\ref{dwdthetaVR=dEperpdthetaVR}). The constant in the integration limits may be arbitrary, as long as the $\chi(E_{\perp})$ dependence is periodic with the period $\Delta E_{\perp}$, and the integration is carried out over the full period.

\begin{figure}
\includegraphics{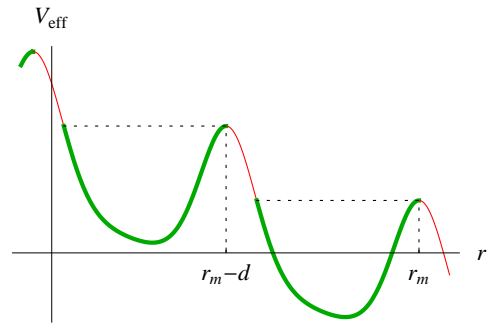}
\caption{\label{fig:red-green-regions} Solid curve, effective potential $V_{\text{eff}}(r)$ of a bent crystal. Its red regions signify loci of radial reflection points of VR particles coming from and returning to $r\to+\infty$ with a constant $E_\perp$. 
Red curve segments combined with the horizontal dotted lines show behaviour of function $\underset{r'\geq r}\max V_{\text{eff}}(r')$.
Thick green curve segments mark intervals contributing to the mean VR angle. }
\end{figure}

Although VR is supposed to be formed deeply inside the bent crystal, in principle, integral (\ref{eq2ab}) has yet some residual dependence on its endpoints. The latter, however, fades away with the increase of the target thickness. If, to the leading order in thickness, this dependence is neglected, there remains the genuine volume contribution. Formally, it expresses as a thick-crystal limit:
\begin{equation}\label{thetaVR=lim}
    \chi
= 2 \underset{r_0\to\infty}\lim  \left\{ \sqrt{2\frac{r_0}{R}}
-\frac1R\int_{r_{\min}(E_{\perp})}^{r_0}dr{\sqrt{\frac{E}{2[E_{\perp}-V_{\text{eff}}(r)]}}}\right\}.
\end{equation}
Here 
\begin{equation}\label{Veff-def}
V_{\text{eff}}(r)=V(r)-\frac{Er}{R}
\end{equation}
is the effective potential of the bent crystal, which has a ``washboard'' shape due to a tilt introduced by the linearized centrifugal potential component (see Fig. \ref{fig:red-green-regions}).
The difference in the braces in Eq. (\ref{thetaVR=lim}) converges on a transverse spatial scale $\Delta r\sim {RV_0}/{E}={R\theta_c^2}/{2}$ corresponding to 
\begin{equation}\label{Delta-z-VR}
\Delta z\sim \sqrt{2R\Delta r}\sim R\theta_c.
\end{equation}
This longitudinal scale must be smaller than the crystal thickness $L=R\theta_b$, where $\theta_b$ is the crystal bending angle:
\begin{equation}\label{L>>Deltaz}
L\gg \Delta z,\quad \text{i.e.}, \quad \theta_b\gg\theta_c.
\end{equation}
The crystals for VR are usually designed to meet the latter requirement. 
By virtue of the mentioned convergence at large $r_0$, Eq. (\ref{thetaVR=lim}) holds equally well for cases when the VR region is not strictly in the middle of the crystal.

For a periodic $V(r)$, the integrand in (\ref{thetaVR=lim}) is not periodic in $r$. But granted the simplicity of its dependence on the period number, it is feasible to sum over the entire sequence of periods in a closed form \cite{Doct-diss}.
That leads to a generic representation for the VR angle as an integral over a single interplanar period with the integrand given by a generalized Riemann zeta-function \cite{ref:zeta} (see also \cite{Bond-VR-PRA,Bond-PRST}) with parameter $1/2$ and the argument depending on $V_{\text{eff}}(r)$:
\begin{equation}\label{theta-VR-zeta}
    \chi
= -\frac{2}{\sqrt{2Rd}}\int_{0}^{d}dr\mathfrak{Re}
\zeta\left(\frac12,\frac{R}{d}\frac{E_{\perp}-V_{\text{eff}}(r)}{E}\right).
\end{equation}
Substitution thereof to Eq. (\ref{const+DeltaEperp}) leads to further simplifications, and yields an expression for the mean VR angle involving no special functions:
\begin{equation}\label{general-mean-thetaVR-max}
\sqrt{\frac{E}{2}}\frac{d}{2}\overline{\chi}(E,R) =\int_0^{d}dr \sqrt{\underset{r'\geq r}\max V_{\text{eff}}(r') -V_{\text{eff}}(r)},
\end{equation}
where $\underset{r'\geq r}\max V_{\text{eff}}(r')$ is the global maximum in the semi-infinite region $r\leq r'<\infty$, thus being unique (see Fig. \ref{fig:red-green-regions}), in contrast to possible minor local maxima.
Formula (\ref{general-mean-thetaVR-max}) can also be derived directly, bypassing evaluation of the angular (or $E_\perp$-) distribution, as is demonstrated in Appendix \ref{app:sqrt-maxV}.

The integrand in Eq.~(\ref{general-mean-thetaVR-max}) differs from zero only in regions $V_{\text{eff}}(r)<\underset{r'\geq r}\max V_{\text{eff}}(r')$, marked in Fig. \ref{fig:red-green-regions} by green. 
Paradoxically, those are the regions usually associated with channeled particles, whereas VR particles, being over-barrier, on the contrary, radially reflect in the regions marked in Fig. \ref{fig:red-green-regions} by red. Nonetheless, the latter regions do not contribute to the mean VR angle at all. 
One of the consequences is that  $\overline{\chi}$ must strictly vanish for $R\leq R_c$ (where $R_c$ by definition is the smallest value of $R$, at which $V_{\text{eff}}(r)$ has local minima, and hence maxima), 
insofar as in that case $V_{\text{eff}}(r)$ has no maxima. 
I.e., VR exists under the same condition
\begin{equation}\label{meanVRneq0}
R> R_c
\end{equation}
as channeling in a bent crystal, but pertains to overbarrier particles. 
That agrees with the former results of numerical evaluation of integral (\ref{Maisheev-repr}) in \cite{Bellucci-Chesnokov-Maisheev-Yazynin} (see also \cite{Breese-Biryukov}).
It is also evident that the integrand of (\ref{general-mean-thetaVR-max}) is everywhere positive, so the net deflection always proceeds to the side opposite to that of the crystal bending. 
As was mentioned in the Introduction, that is the distinguishing feature of VR.

In a periodic continuous potential, all of whose wells are equivalent [such as the potential of a silicon crystal in orientation (110)], it is convenient to choose the integration interval end point coinciding with location of any of the maxima $r_m$ of the effective potential: $\underset{r'\geq r}\max V_{\text{eff}}(r')= V_{\text{eff}}(r_m)$. 
For such a choice, $\underset{r'\geq r}\max V_{\text{eff}}(r')$ does not depend on $r$ within the entire part of integration interval, where the radicand is greater than zero. Moreover, the integration may be extended over the entire crystal period, provided the region where the radicand turns negative (and the integrand imaginary) is  eliminated by taking the real part:
\begin{equation}\label{general-mean-thetaVR}
\sqrt{\frac{E}{2}}\frac{d}{2}\overline{\chi}(E,R) 
=\int_{r_{m}-d}^{r_{m}}dr\mathfrak{Re}\sqrt{ V_{\text{eff}}(r_{m}) -V_{\text{eff}}(r)}.
\end{equation}
That obviates the need for independent maximization of $V_{\text{eff}}$ for every value of $r$ in the integrand.

In a more complicated case, when the continuous potential contains two inequivalent wells per period, as it is known to be, e.g., for positively charged particles in a silicon crystal in orientation (111) (see Fig. \ref{fig:contin-pot}b below), it suffices to split the integration interval in two parts by the intermediate maximum:
\begin{eqnarray}\label{general-mean-thetaVR-two-wells}
\sqrt{\frac{E}{2}}\frac{d}{2}\overline{\chi}(E,R) 
=\int_{r_{m2}-d}^{r_{m1}}dr\mathfrak{Re}\sqrt{ V_{\text{eff}}(r_{m1}) -V_{\text{eff}}(r)}\nonumber\\
+\int_{r_{m1}}^{r_{m2}}dr\mathfrak{Re}\sqrt{V_{\text{eff}}(r_{m2}) -V_{\text{eff}}(r)}.\quad
\end{eqnarray}
The subsequent calculation procedure for each of the latter partial integrals is the same as for integral (\ref{general-mean-thetaVR}) for a single-well potential.

It is also possible [e.g., for negatively charged particles and crystal orientation (111), corresponding to Fig. \ref{fig:contin-pot}b flipped upside down] that at small $E/R$ the potential is of single-well type, while with the increase of $E/R$ it becomes double-well. In that case, at the critical $E/R$ value one must switch from formula (\ref{general-mean-thetaVR}) to (\ref{general-mean-thetaVR-two-wells}).

Formulas (\ref{general-mean-thetaVR}), (\ref{general-mean-thetaVR-two-wells}) are well suited both for numerical evaluation of the mean VR angle and for analysis of the intrinsic $E$- and $R$-dependencies. The latter will be the subject of the next three sections. Phenomenological issues are discussed in Sec. \ref{sec:phenomenology}.

\section{Small-$R_c/R$ expansion}\label{sec:expansion-RcR}

For a given oriented crystal, i.e., given $V(r)$, the rhs of Eq. (\ref{general-mean-thetaVR-max}) depends on the experimentally changeable parameters $E$ and $R$ only via the ratio $E/R$, representing the centrifugal force. 
For any shape of the interplanar potential, the increase of centrifugal force $E/R$ makes the effective potential well progressively more tilted and shallower, wherewith the mean reflection angle (\ref{general-mean-thetaVR}) decreases. 
Vice versa, its maximal value is achieved in the formal straight-crystal limit\footnote{\label{foot:E/R->0}Throughout this paper, we assume the incoherent multiple scattering effect on $\overline{\chi}$ to be negligible, despite the expanding intrinsic VR region (\ref{Delta-z-VR}) -- see condition (\ref{RllRmult}), (\ref{Rmult-def}) below. The mentioned limit thus must be understood in the sense of extrapolation. Since $R$ is a dimensional quantity, condition $R\to\infty$ musty be understood in the present context as $4R_c/R\ll1$ or $\frac{Ed}{V_0R}\ll1$.} $R\to\infty$. 
That is the second salient feature of VR mentioned in the Introduction. Since the case of small-$R_c/R$, when $\overline{\chi}$ is maximal, is of the highest practical value, we will analyse it in the first place. 

\subsection{Leading order in $R_c/R$}

To evaluate the maximal value of $\overline{\chi}$ achieved at $R\to\infty$, one needs merely to replace in Eq. (\ref{general-mean-thetaVR-max}) the effective potential by the real one:
\begin{equation}\label{mean-thetaVR-infty}
\underset{R\to\infty}\lim\overline{\chi}=\chi_0=\sqrt{\frac{2}{E}}\frac{2}{d}\int_0^{d}dr\sqrt{\max V -V(r)}.
\end{equation}
Since the integrand in the right-hand side of Eq. (\ref{mean-thetaVR-infty}) is a periodic function of $r$, it does not matter where to choose the integration limits of an interval of the length $d$.

The obtained result can be cast in a more intuitive form
\begin{equation}\label{2localthetac}
\chi_0=2\left\langle \theta_c(r) \right\rangle_{r},
\end{equation}
where 
\begin{equation}\label{thetac(r)-def}
\theta_c(r)=\sqrt{2\left[\max V -V(r)\right]/E}
\end{equation}
is the `local' critical angle (for a straight crystal) sensed by the overbarrier particle during its passage through the interplanar interval with a nearly critical transverse energy, and $\left\langle ... \right\rangle_{r}=\frac1d\int_0^{d}dr ...$ designates averaging over the continuous potential period. The factor of 2 in Eq.~(\ref{2localthetac}) may be interpreted as accounting for 
contributions from particle motion before and after the reflection point. 
A corollary from Eq. (\ref{2localthetac}) is that 
\begin{equation}\label{thetaVR-upperbound}
\chi_0\leq 2\max \theta_c(r) \equiv2\theta_c.
\end{equation}

However, in practice, $R$ is never so large compared to $R_c$ that the difference between $\overline{\chi}$ and $\chi_0$ is really negligible. Thus, at least a first-order $E/R$-dependent correction to $\chi_0$ should be taken into account. Its derivation and refinements are discussed below. 
% is detailed below

\subsection{Next-to-leading order in $R_c/R$}

Since Eq. (\ref{general-mean-thetaVR-two-wells}) reduces the problem for a two-well periodic continuous potential to that for a single-well one, we begin with the most elementary case when the generic solution is given by formula (\ref{general-mean-thetaVR}).

\begin{figure}
\includegraphics{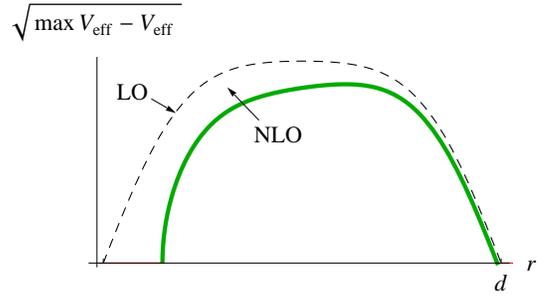}
\caption{\label{fig:LO-NLO} Solid curve, the integrand of Eq. (\ref{general-mean-thetaVR}) for a tilted single-well periodic potential. Dashed curve, the $R_c/R\to0$ limit of the integrand. The area under it is the LO contribution [Eq. (\ref{mean-thetaVR-infty})]. The small area between the solid and the dashed curves at small $R_c/R$ may be regarded as the NLO contribution. }
\end{figure}

To derive from it the next-to-leading order (NLO) correction to (\ref{mean-thetaVR-infty}), 
note that in Eq. (\ref{general-mean-thetaVR}) the dependence on $E/R$ enters both to $ V_{\text{eff}}(r)$ and $ V_{\text{eff}}(r_m)$, where $r_m$ depends on $E/R$, too. 
If $V(r)$ has maxima at $r=0$ and $r=d$, the effective potential maximum location $r_m$ may be chosen to be close to $d$. In its vicinity,
\begin{eqnarray}
V_{\text{eff}}(r)&\underset{r\approx d}\simeq&\max V-\frac{(r-d)^2}{2}|V''(d)|-\frac{E}{R}r\nonumber\\
&=&\max V-\frac{Ed}{R}-\frac{(r-r_m)^2}{2}|V''(d)|+\mathcal{O}\left(\frac{E^2}{R^2}\right),\nonumber\\
\end{eqnarray}
where we took into account that $V'(d)=0$. Hence, the potential maximum location shifted by the centrifugal tilt equals
\[
r_m=d-\frac{E}{R|V''(d)|}+\mathcal{O}\left(\frac{E^2}{R^2}\right).
\]
This allows to determine the corresponding effective potential maximal value.
\[
V_{\text{eff}}(r_m)=\max V-\frac{Ed}{R}+\mathcal{O}\left(\frac{E^2}{R^2}\right),
\]
which proves to be independent of $V''(d)$ in this approximation.
Therefore, to the NLO, i.e., omitting all the $\mathcal{O}\left({E^2}/{R^2}\right)$ contributions, all the linear $E/R$ dependence in (\ref{general-mean-thetaVR}) reduces to a term $\frac{E}{R}(r-d)$ in the radicand:
\begin{eqnarray}\label{linearER-under-sqrt}
\sqrt{\frac{E}{2}}\frac{d}{2}\overline{\chi}=\mathfrak{Re}\int_0^{d}dr\sqrt{\max V-V(r)+\frac{E}{R}(r-d)}\nonumber\\
+\mathcal{O}\left(\frac{E^2}{R^2}\right).\qquad
\end{eqnarray}
The change in the integrand compared with the leading-order (LO) Eq. (\ref{mean-thetaVR-infty}) is illustrated in Fig. \ref{fig:LO-NLO}.

Next, linearizing the entire integrand in $E/R$ and integrating the result termwise, we get
\begin{eqnarray}\label{intsqrtV+int1sqrtV}
\sqrt{\frac{E}{2}}\frac{d}{2}\overline{\chi} \simeq \int_{0}^{d}dr\sqrt{\max V -V(r)}\qquad\nonumber\\
+\frac{E}{2R}\int_{0}^{d}dr\frac{r-d}{\sqrt{\max V -V(r)}}.
\end{eqnarray}
It involves only the potential for a straight crystal. 
Since a single-well potential of a straight crystal must be symmetric wrt its minimum at $r=d/2$, Eq. (\ref{intsqrtV+int1sqrtV}) may be recast as
\begin{subequations}\label{mean-thetaVR-NLO-sharpedge-ab}
\begin{eqnarray}
\overline{\chi} &\simeq&\sqrt{\frac{2}{E}}\frac{4}{d}\int_0^{d/2}dr\sqrt{\max V -V(r)}\nonumber\\
&\,&-\frac{\sqrt{2E}}{R}\int_0^{d/2}\frac{dr}{\sqrt{\max V -V(r)}}\label{mean-thetaVR-NLO-sharpedge}\\
&\equiv&\chi_0-\frac{d}{R}\left\langle\frac1{\theta_c(r)}\right\rangle_r.
\end{eqnarray}
\end{subequations}

%in the correction integral in (\ref{mean-thetaVR-NLO-sharpedge}), due to the periodicity of its integrands, the integration limits were written simply as 0 and $d$. 

Eq. (\ref{mean-thetaVR-NLO-sharpedge}) demonstrates that the deeper on the average the potential well, or, for a fixed depth, the more rectangular it is, the larger is the LO contribution (the first term), and the smaller by absolute value the negative NLO contribution (the second term). Hence, larger VR angles (beneficial for beam steering applications) are achieved for crystal materials with higher $Z$ and orientations with wider interplanar intervals.

To invariantly relate the $\overline{\chi}(E,R)$ dependence in the NLO with the shape of the potential well, one can introduce a product
\begin{subequations}
\begin{eqnarray}
K&=&\frac{\sqrt{E}}{2d}\chi_0\left|\frac{d(\sqrt{E}\overline{\chi})}{d(E/R)}\right|_{E/R\to0}
=\frac{\chi_0}{2d}\left|\frac{\partial\overline{\chi}}{\partial(1/R)}\right|_{R\to\infty}\nonumber\\
%\left\langle\theta_c(r)\right\rangle_r\left\langle\frac1{\theta_c(r)}\right\rangle_r-1
\label{K-def}\\
&=&\left\langle\theta_c(r)\right\rangle_r \left\langle\frac1{\theta_c(r)}\right\rangle_r
\qquad\qquad\qquad\qquad\qquad\qquad\nonumber\\
&\equiv&\left\langle\sqrt{\max V-V(r)}\right\rangle_r \left\langle\frac1{\sqrt{\max V-V(r)}}\right\rangle_r,\qquad\label{subtracted-product}
\end{eqnarray}
\end{subequations}
which, according to the Cauchy-Bunyakovsky-Schwarz inequality, must be greater than unity: 
\begin{equation}\label{Kgeq1}
K\geq1
\end{equation}
for any $V(r)$. The smaller (the closer to unity) this number, the closer the well shape to rectangular.

It should be minded, however, that Eq. (\ref{subtracted-product}) and inequality (\ref{Kgeq1}) have been  established only for a single-well potential, and only provided $\int_0^{d/2}\frac{dr}{\sqrt{\max V -V(r)}}$ converges, by virtue of which the integration may be extended over the full interplanar interval. Otherwise, notably, for multi-well potentials, the rhs of (\ref{K-def}) may become smaller than unity, because in the limit $R_c/R\to0$ not all the particles completely traverse the last interplanar interval, being reflected in different sub-wells (see examples in Sec. \ref{sec:expansion-RcR} below).

For a double-well periodic potential, a similar analysis can be based on Eq. (\ref{general-mean-thetaVR-two-wells}). Repeating the linearisation procedure for each of the intervals, and presuming the symmetry of each sub-well (which must hold since inversion wrt each sub-well center leaves the periodic sequence of atomic planes invariant),
one is led to the result
\begin{subequations}
\begin{eqnarray}\label{}
\overline{\chi}&\simeq&\chi_0-\frac{\sqrt{2E}}{R}\Bigg(\frac{d_1}{d_1+d_2}\int_{0}^{d_1/2}\frac{dr}{\sqrt{\max V -V(r)}} \nonumber\\
&\,&\qquad +\frac{d_2}{d_1+d_2}\int_{d_1}^{d_1+d_2/2}\frac{dr}{\sqrt{\max V -V(r)}}\Bigg)\qquad\\
&=&\frac{2}{d}\left(d_1\left\langle\theta_c\right\rangle_1+d_2\left\langle\theta_c\right\rangle_2\right)\nonumber\\
&\,&\qquad -\frac{1}{Rd}\left(d_1^2\left\langle\frac{1}{\theta_c}\right\rangle_1+d_2^2\left\langle\frac{1}{\theta_c}\right\rangle_2\right),
\end{eqnarray}
\end{subequations}
where $\left\langle \ldots \right\rangle_1=\frac{1}{d_1}\int_0^{d_1}dr\ldots$, $\left\langle \ldots \right\rangle_2=\frac{1}{d_2}\int_{d_1}^{d}dr\ldots$.
For $d_1=0$ or $d_2=0$, as well as for $d_1=d_2=d/2$, this goes over to Eq. (\ref{mean-thetaVR-NLO-sharpedge-ab}).

\subsection{Logarithmic modification of NLO term}\label{subsec:NLO-log-modif}

Unfortunately, the applicability of simple formula (\ref{mean-thetaVR-NLO-sharpedge}) in practice is undermined because the difference $\max V -V(r)$ at the tops of the barriers tends to zero quadratically, wherewith $\int_0^{d/2}\frac{dr}{\sqrt{\max V -V(r)}}$ diverges logarithmically. For positively charged particles that happens due to thermal (and zero-point) fluctuations of atom positions in the planes. 
Granted that those fluctuations are relatively small, a finite result might be attained by just replacing in the NLO integral in Eq.~(\ref{mean-thetaVR-NLO-sharpedge}) the interplanar potential by its static counterpart [i.e., letting $u=0$ in Eq. (\ref{VP=conv-Gauss-VPS}) below]:
\begin{equation}\label{NLO-Vstatic}
\overline{\chi} \simeq\chi_0-\frac{\sqrt{2E}}{R}\int_0^{d/2}\frac{dr}{\sqrt{\max V_{\text{stat}} -V_{\text{stat}}(r)}},
\end{equation}
with
\[
V_{\text{stat}}(r)=V(r)\big|_{u=0}.
\]
Such an approximation should not be too unreasonable, because if the radial reflection point belongs to the region $r_c>u$, for all $r>r_c>u$ the difference between $\max V_{\text{stat}} -V_{\text{stat}}(r)$ and $\max V -V(r)$ is small. 

Alternatively, one may introduce in (\ref{mean-thetaVR-NLO-sharpedge}) an appropriate logarithmic cutoff $r_*\ll d/2$:
\begin{equation}\label{NLO-cutoff}
\overline{\chi} \simeq \chi_0
-\frac{\sqrt{2E}}{R}\int_{r_*}^{d/2}\frac{dr}{\sqrt{\max V -V(r)}}.
\end{equation}
Here, assuming the symmetry of the interplanar potential, we integrated only over half the period and multiplied by 2. 
%and placed the maximum of $V(r)$ at the origin, $\max V=V(0)$.
The error due to the deliberateness of the choice of $r_*$ may be relatively small. 

For negatively charged particles, however, such an ad hoc cutoff approach may be too crude, because the round top of the potential barrier lies between the planes, being broad for any crystal temperature. 
In that case, to be rigorous, one should return to representation (\ref{linearER-under-sqrt}). 
It can be handled in the next-to-leading logarithmic order by splitting the integration interval in two parts, in the first of which, $0<r<r_i\ll d$, the potential is approximable by its Taylor expansion around the maximum up to the quadratic term: $V(r)\simeq\max V+V''(0)r^2/2$, with $V''(0)<0$, whereas in the second, $r_i<r<d$, it is already safe to expand the radicand to the NLO in $R_c/R$, as in Eq. (\ref{intsqrtV+int1sqrtV}):
\begin{eqnarray}\label{split-r0}
\sqrt{\frac{E}{2}}\frac{d}{2}\overline{\chi}\simeq\mathfrak{Re}\int_0^{r_i}dr\sqrt{ \frac{|V''(0)|}{2}r^2-\frac{Ed}{R}}\qquad\qquad\quad\nonumber\\
+\int_{r_i}^{d}dr\left[\sqrt{\max V-V(r)}+\frac{E (r-d)}{2R\sqrt{\max V-V(r)}}\right]\nonumber\\
+\mathcal{O}\left(\frac{E^2}{R^2}\right).\quad
\end{eqnarray}
Computing those two integrals and eliminating $r_i$ (see Appendix \ref{app:NLO-log}), we are led to a form similar to (\ref{NLO-cutoff}),
\begin{equation}\label{NLOthetaVR-r*(ER)}
\overline{\chi} \simeq\chi_0-\frac{\sqrt{2E}}{R}\int_{r_*(E/R)}^{d/2}\frac{dr}{\sqrt{\max V -V(r)}},
\end{equation}
with the difference  that the lower cutoff is now unambiguously defined and centrifugal-force-dependent:
\begin{equation}\label{r*-def}
r_*(E/R)=\sqrt{\frac{Ed}{2eR|V''(0)|}},
\end{equation}
where $e=2.718$ is the base of a natural logarithm.

The obtained value of $r_*$ is natural by the order of magnitude, because the corresponding drop of the potential near the top is commensurable with a typical centrifugal energy
\[
|V''(0)|\frac{r_*^2}{2}=\frac{Ed}{4eR}\sim \Delta E_{\perp}.
\]
Noteworthy, however, is a small numerical factor $\frac{1}{4e}\approx 0.1$. It corroborates our assumption that $r_*\ll d/2$, as long as
\[
R_c=\frac{E}{F_{\max}}<\frac{2E}{|V''(0)|d}=\left(\frac{2r_*}{d}\right)^2eR\ll R.
\]

Thus, in the case when the particle reflects from the potential barrier in the region of its parabolic top, 
for description of the NLO correction it suffices to know only two empirical constants: $V''(0)$ and
\begin{eqnarray}
\underset{r_*\to0}\lim \left(\int_{r_*}^{d/2}\frac{dr}{\sqrt{\max V -V(r)}}+\sqrt{\frac{2}{|V''(0)|}}\ln\frac{2r_*}{d}\right)\nonumber\\
=\int_0^{d/2}dr\ln\frac{d}{2r}\frac{d}{dr}\frac{r}{\sqrt{\max V-V(r)}}\quad
\end{eqnarray}
[see Eq. (\ref{NLO-log-general})]. They can be computed numerically, given a realistic parametrization for the interplanar potential $V(r)$. 

Formula (\ref{NLOthetaVR-r*(ER)}) with cutoff (\ref{r*-def}) is best suited for negatively charged particles, for which the round top of the potential barrier is wide, wherewith the Taylor expansion around its maximum works in a sufficiently broad interval of $E/R$, too. 
%$|\kappa|\ll1$
%granted that $\frac{d}{dr}\frac{r}{\sqrt{\max V-V(r)}}$ is small. 
%Then little depends on the detail of the shape of the potential well, except on $V(0)$ for $\chi_0$ and $V''(0)$ for the slope. In particular, the temperature dependence is weak.
As for positively charged particles, for them
the condition implied at derivation of Eq. (\ref{NLOthetaVR-r*(ER)}) may be violated if the tilt proportional to $E/R$ is  so large that in the reflection point the continuous potential top is no longer parabolic. But even in that case, around the turnover point the distinction from the extended parabolic approximation for the potential top may be relatively mild (otherwise the NLO approximation itself can break down). 
Whether or not the logarithmic dependence along with cutoff prescription (\ref{r*-def}) survive under such conditions will be investigated in more detail on model examples in Sec. \ref{subsubsec:110poschar} and for realistic potentials in Sec. \ref{sec:phenomenology}.
%Since for positively charged particles $|V''(0)|$ is large, it holds that $\kappa\gg1$, so formula (\ref{NLO-log-general}) may merely be viewed as an analogue of Eq. (\ref{NLO-cutoff}), with  $r_*$.

%For a purely parabolic potential for negatively charged particles, 
%\begin{eqnarray}
%\sqrt{\frac{E}{2}}\overline{\chi}
%=\sqrt{V_0}-\frac{Ed}{4R\sqrt{V_0}}\left(1+\ln\frac{4V_0R}{Ed}\right)+\mathcal{O}\left[\left(\frac{E}{R}\right)^2\right].
%\end{eqnarray}

If the potential has two inequivalent wells per period, with widths $d_1$ and $d_2$, the derivation of the NLO formula must be based on Eq. (\ref{general-mean-thetaVR-two-wells}). It gives
\begin{eqnarray}\label{}
\overline{\chi}\simeq\chi_0-\frac{\sqrt{2E}}{R}\Bigg(\frac{d_1}{d_1+d_2}\int_{r_{*1}}^{d_1/2}\frac{dr}{\sqrt{\max V -V(r)}} \nonumber\\
+\frac{d_2}{d_1+d_2}\int_{d_1+r_{*2}}^{d_1+d_2/2}\frac{dr}{\sqrt{\max V -V(r)}}\Bigg),\quad
\end{eqnarray}
\[
r_{*k}(E/R)=\sqrt{\frac{Ed_k}{2eR|V''(r_{mk})|}}, \qquad k=1\text{ or }2.
\]

\section{Model results}\label{sec:models}

Even though the obtained generic solution (\ref{general-mean-thetaVR}) or (\ref{general-mean-thetaVR-two-wells}) reduces the problem to an integral, which is sufficiently simply calculable numerically, it can be useful sometimes to refer also to model results, 
in which the dependencies of $\overline{\chi}$ on all the parameters are explicit. 
Models can also be used for testing the accuracy of the proposed generic approximations, such as the NLO expansion. They may be even more helpful when $R$ becomes commensurable with the critical value $R_c$, so that the NLO approximation breaks down. 
But it is desirable in this case that the models reflect the shape of the interplanar potential adequately enough. In this section, for references, we will quote predictions for $\overline{\chi}$ for  a few such model potentials, pertinent both to single-well and double-well cases.

\subsection{Simple parabolic and square well potentials. Arbitrary $R/R_c$}\label{subsec:SimplestModels}

Integral (\ref{general-mean-thetaVR}) can be taken in elementary functions only if $V(r)$ is a linear or quadratic polynomial in $r$ (with an exception considered in Sec. \ref{sec:Rtorc}). 
%It may seem that on the entire interplanar interval the potential can hardly be realistically approximated by a single parabola or a square well. 
In fact, for silicon crystal in orientation (110) the parabolic approximation is known to be rather satisfactory. It is this approximation that was used in \cite{Bond-VR-PRA}, but only for small $R/R_c$. 
More generally, if the effective potential is strongly tilted, only relatively small portions of the wells contribute to the mean VR angle (cf. Fig. \ref{fig:red-green-regions}). 
Then, if the potential varies smoothly enough, as is typical for moderate-$Z$ crystal materials, it can be parametrized in those small regions by a quadratic polynomial. On the other hand, if the potential walls are steep, as is typical for high-$Z$ materials, one can approximate them by square wells. It will be instructive first to analyse properties of those two simplest among exactly solvable models.

\begin{itemize}
\item
For a sequence of parabolic wells of depth $V_0>0$, described within a period by the potential 
%\begin{subequations}\label{mean-thetaVR-harm-positive-110ab}
\begin{equation}\label{V-pos-pure-parab}
V(r)=V_0\left(\frac{2r}{d}-1\right)^2, \qquad 0<r<d,
\end{equation}
serving as an idealization for the interplanar potential for positively charged particles in a silicon crystal in orientation (110), evaluation of integral (\ref{general-mean-thetaVR}) gives
%\begin{subequations}\label{mean-thetaVR-harm-positive-110ab}
\begin{equation}\label{mean-thetaVR-harm-positive-110}
    \frac{\overline{\chi}}{\theta_c}=\frac{\pi}{2} \left(1-\frac{R_c}{R}\right)^2
\end{equation}
with 
\begin{equation}\label{Rc-parab}
R_c=\frac{E}{F_{\max}}=\frac{Ed}{4V_0}
\end{equation}
being the critical radius for this model.
In the LO, $\chi_0=\frac{\pi}{2}\theta_c$, while in the NLO, the squared binomial in Eq. (\ref{mean-thetaVR-harm-positive-110}) linearizes to
\begin{equation}\label{mean-thetaVR-harm-positive-110-NLO}
    \frac{\overline{\chi}}{\theta_c}=\frac{\pi}{2} \left(1-\frac{2R_c}{R}\right)+\mathcal{O}\left(\frac{R_c^2}{R^2}\right),
\end{equation}
%\end{subequations}
agreeing with the result of \cite{Bond-VR-PRA}, as well as with the generic formula (\ref{mean-thetaVR-NLO-sharpedge}).

Dimensionless parameter (\ref{subtracted-product}) for this model equals
\begin{equation}\label{pi28}
K=\frac{\pi^2}{8}=1.234.
\end{equation}

\begin{figure}
\includegraphics{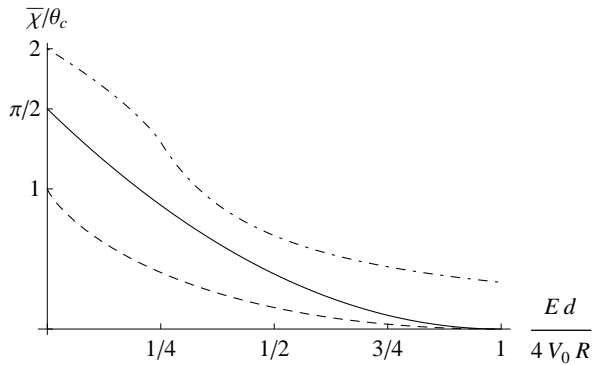}
\caption{\label{fig:meanVR-parab-squarewell} Mean VR angle dependence on $E/R$ for several models for the periodic continuous potential with well depth $V_0$ and width $d$. Solid parabola, positively charged particles in a harmonic potential well [Eq. (\ref{mean-thetaVR-harm-positive-110})]. Dashed curve, negatively charged particles in the same field [Eq. (\ref{mean-thetaVR-harm-negative-110})]. Dot-dashed curve, the result for a square well interplanar potential, Eq. (\ref{mean-thetaVR-squarewell}). 
Concerning the limit $E/R\to0$, see footnotes \ref{foot:Ev2}, \ref{foot:E/R->0}.}
\end{figure}

\item
For the sequence of parabolic barriers of height $V_0>0$, being an idealization for the continuous potential for negatively charged particles in silicon in orientation (110), the potential well has the form
\[
V(r)=\begin{cases}
-V_0\left(\frac{2r}{d}\right)^2,  & 0<r<d/2,\\
-V_0\left(2\frac{d-r}{d}\right)^2,  & d/2<r<d.
\end{cases}
\]
Inserting this to Eq. (\ref{general-mean-thetaVR}), one obtains
\begin{equation}\label{mean-thetaVR-harm-negative-110}
    \frac{\overline{\chi}}{\theta_c}=
1-\frac{R_c}{R} \left(1+\ln\frac{R}{R_c}\right),
\end{equation}
where $R_c$ is given by the same Eq. (\ref{Rc-parab}). For this model, $\chi_0=\theta_c$, while $\frac{\chi_0}{2d}\big|\frac{\partial\overline{\chi}}{\partial(1/R)}\big|$ at $R_c/R\to0$ retains a logarithmic dependence on $R_c/R$. Right-hand sides of both (\ref{mean-thetaVR-harm-positive-110}) and (\ref{mean-thetaVR-harm-negative-110}) at $R\to R_c$ tend to zero \emph{quadratically}, because at $R\to R_c$ in integral (\ref{general-mean-thetaVR}) both the width and the height of the integrand $\mathfrak{Re}\sqrt{\max V_{\text{eff}}-V_{\text{eff}}(r)}$ tend to zero linearly. 
Formula (\ref{mean-thetaVR-harm-negative-110}) is exact, but follows as well from the generic NLO approximation (\ref{NLOthetaVR-r*(ER)}) when setting there $V(r)\simeq\frac12V''(0)r^2$ with $V''(0)=-8V_0/d^2$. All the corrections beyond the NLO in this model happen to vanish.

\item
In the opposite extreme of a square well, the potential within the period has a constant depth $V_0>0$, viz., $V(r)=-V_0$ within the interval $0<r<d$, and $V=0$ at $r=0$ and $r=d$.\footnote{It represents a zero-barrier-thickness limit of the model considered in \cite{Kovalev}. This may be regarded as a $Z\to \infty$ limit for positively charged particles in a crystal in orientation (110), although  even for tungsten, the well shape is far from rectangular yet. For negatively charged particles in such a potential, $\overline{\chi}$ would equal zero.} 
Then integral (\ref{general-mean-thetaVR}) evaluates to
\begin{equation}\label{mean-thetaVR-squarewell}
    \frac{\overline{\chi}}{\theta_c}=\frac{4RV_0}{3Ed}\left[1-\mathfrak{Re}\left(1-\frac{Ed}{RV_0}\right)^{3/2}\right],
\end{equation}
which for $\frac{Ed}{RV_0}<1$ matches with the result of \cite{Kovalev} in the limit of zero barrier thickness ($a\to0$).
The rhs of this expression as a function of $R$ tends to zero (linearly) only at $R\to 0$, because 
for a square well $R_c=0$. The linear law here reflects the fact that at $R\to R_c$
only the width of the effective potential well shrinks to zero, whereas its depth remains constant, being determined by the height of the sharp walls. 
Three terms of expansion of (\ref{mean-thetaVR-squarewell}) read
\begin{equation}\label{square-well-NLO}
    \frac{\overline{\chi}}{\theta_c}=2-\frac{Ed}{2RV_0}-\frac13 \left(\frac{Ed}{2RV_0}\right)^2+\mathcal{O}\left[\left(\frac{Ed}{RV_0}\right)^3\right].
\end{equation}
So, for this model, $\chi_0=2\theta_c$ [the upper bound of (\ref{thetaVR-upperbound}) is reached], and
\begin{equation}\label{}
%\frac{\chi_0}{2d}\left|\frac{d\overline{\chi}}{d(1/R)}\right|_{E/R\to0}
%=\overline{\chi}\left\langle\theta_c(r)\right\rangle_r\left\langle\frac1{\theta_c(r)}\right\rangle_r
K=1
\end{equation}
[the lower bound of (\ref{Kgeq1}) is reached], because $\theta_c(r)$ here is just a constant.
It is also noteworthy that for a square well, the next-to-next-to-leading order (NNLO) contribution to $\overline{\chi}$ is negative, whereas for parabolic well, according to Eq. (\ref{mean-thetaVR-harm-positive-110}) expanded up to quadratic term in $R_c/R$, it was positive.

\end{itemize}

The behavior of dependencies (\ref{mean-thetaVR-harm-positive-110}), (\ref{mean-thetaVR-harm-negative-110}) and (\ref{mean-thetaVR-squarewell}) is illustrated in Fig. \ref{fig:meanVR-parab-squarewell}.

\subsection{Double-parabolic potential}

More elaborate models may serve to illustrate the dependence of $\chi_0$ and the slope of $\overline{\chi}(E/R)$ on some detail of the shape of the interplanar potential, both for positively and for negatively charged particles.
In principle, the integration can be done exactly for any piecewise-parabolic or piecewise-linear potential. For orientation (110), two parabolas can interpolate the planar potential within a period reasonably enough. 
For orientation (111), where the distances between the planes are not all equal but alternate, producing two inequivalent wells per period (see Fig. \ref{fig:contin-pot}b below), that is necessary even in the crudest approximation. Let us thus consider the latter case first.

\subsubsection{Orientation (111), positively charged particles}\label{subsubsec:111pos}

The continuous potential for positively charged particles in a silicon crystal in orientation (111) features two unequal potential wells per period (see Fig. \ref{fig:contin-pot}b below). For simplicity, let us neglect here the thermal smearing of the continuous potential near the planes, treating the potential as consisting of just two alternating parabolic wells of different widths $d_S$, $d_L$ and depths $V_S$, $V_L>0$. To facilitate correspondence with the cases of negatively charged particles, as well as orientation (110) considered below, it is expedient to write the continuous potential within a period in a symmetric form: 
\begin{equation}\label{V-piecewise-parab-111}
V(r)=\begin{cases}
\mathcal{V}_1(r),  & 0\leq r\leq d_L/2,\\
\mathcal{V}_2(r),  & d_L/2\leq r\leq d-d_L/2,\\
\mathcal{V}_3(r),  & d-d_L/2\leq r\leq d,
\end{cases}
\end{equation} 
where
\begin{equation}\label{}
\mathcal{V}_1(r)=V_L\left(\frac{2r}{d_L}\right)^2
\end{equation}
and
\begin{equation}\label{}
\mathcal{V}_3(r)=\mathcal{V}_1(d-r)=V_L\left(2\frac{d-r}{d_L}\right)^2
\end{equation}
are the halves of the two  deep and broad wells, and 
\begin{equation}\label{}
\mathcal{V}_2(r)=V_L-V_S+V_S\left(\frac{d-2r}{d_S}\right)^2,
\end{equation}
with $d_L<d$, $V_S<V_L$, describes a minor midway well.

Under such conditions, we can apply formula (\ref{general-mean-thetaVR-two-wells}). 
Evaluation of the corresponding partial integrals gives
\begin{equation}\label{mean-thetaVR-harm-positive-111}
    \overline{\chi}=\frac{\pi}{2d} \left[d_L\theta_{cL}\left(1-\frac{R_{cL}}{R}\right)^2
+d_S\theta_{cS}\left(1-\frac{R_{cS}}{R}\right)^2\right],
\end{equation}
with
\[
\theta_{cL}=\sqrt{\frac{2V_L}{E}}, \,\, \theta_{cS}=\sqrt{\frac{2V_S}{E}}, \,\, R_{cL}=\frac{Ed_L}{4V_L}, \,\, R_{cS}=\frac{Ed_S}{4V_S}.
\]
At that,
\begin{equation}\label{}
    \chi_0=\frac{\pi}{2} \left(\frac{d_L}{d}\theta_{cL}
    +\frac{d_S}{d}\theta_{cS}\right).
\end{equation}
Therefore, the mean VR angle (\ref{mean-thetaVR-harm-positive-111}) in this case amounts the average of critical angles for each of the sub-wells [cf.  Eq. (\ref{mean-thetaVR-harm-positive-110})], 
with weights proportional to the widths of the corresponding intervals. For the limiting value $\chi_0$ that is rather obvious, in view of representation (\ref{2localthetac}). Its validity through all orders in $R_c/R$ owes to the similarity between the two unequal wells. 

If condition $V_L/d_L=V_S/d_S$ is satisfied, wherewith $R_{cL}=R_{cS}=R_{c}$, then
\begin{equation}\label{mean-thetaVR-harm-positive-111-Rc1=Rc2}
    \overline{\chi}=\chi_0\left(1-\frac{R_{c}}{R}\right)^2
\end{equation}
[with $R$-dependence being similar to that in Eq. (\ref{mean-thetaVR-harm-positive-110})]. 
Taking into account that for a real silicon crystal in (111) orientation $d_L/d_S=3$, we find
\begin{equation}\label{thetaVRinftypos111}
    \chi_0=\frac{\pi\theta_{cL}}{8}\left(3+3^{-1/2}\right)
   =1.4\,\theta_c.
\end{equation}
At that, the product defined by Eq. (\ref{K-def}) equals
\begin{equation}\label{}
%\frac{\chi_0}{2d}\left|\frac{\partial\overline{\chi}}{\partial(1/R)}\right|_{E/R\to0}
K=\frac{3\pi^2}{2^9}\left(3+3^{-1/2}\right)^2=0.74<1.
\end{equation}
It illustrates that inequality (\ref{Kgeq1}) may break down for a double well.

\subsubsection{Orientation (111), negatively charged particles}\label{subsubsec:111neg}

For negatively charged particles in the same crystal orientation (111), the continuous potential is inverted upside down. Then, in effect, it contains just one well, which merely features a small bump in the middle of its bottom. For the present case, calculation by exact formula (\ref{general-mean-thetaVR}) leads to a rather bulky result, so, for simplicity, we will restrict ourselves to the NLO calculation.

The corresponding limiting angle (\ref{mean-thetaVR-infty}) is computed as 
\begin{eqnarray}\label{thetaVR111-arth}
\chi_0&=&\sqrt{\frac{2}{E}}\frac{4}{d}\Bigg[\int_0^{d_L/2} dr\sqrt{\mathcal{V}_1(r)}
%\nonumber\\ &\,&\qquad\qquad
 +\int_{d_L/2}^{d/2}dr\sqrt{\mathcal{V}_2(r)}\Bigg]
\nonumber\\
&=&
\sqrt{\frac{2V_L}{E}}+\frac{d_S}{d}\sqrt{\frac{2V_S}{E}}\left(\frac{V_L}{V_S}-1\right)\text{arth}\sqrt{\frac{V_S}{V_L}}.
\end{eqnarray}
If $V_S<V_L/2$, arth in Eq. (\ref{thetaVR111-arth}) may be expanded as $\text{arth}\sqrt{\xi}\underset{\xi<1/2}\simeq\sqrt{\xi}(1+\xi/3)$, leaving
\begin{equation}\label{}
\chi_0\underset{V_S<V_L/2}\simeq\sqrt{\frac{2V_L}{E}}\left[1+\frac{d_S}{d}\left(1-\frac{2V_S}{3V_L}\right)\right].
\end{equation}
Putting there values $d_S/d=1/4$ and $V_S/V_L\approx 1/3$ corresponding to a real silicon crystal, one finds 
\begin{equation}\label{}
\chi_0=\sqrt{\frac{2V_L}{E}}\left(1+\frac{7}{36}\right)\approx1.2\,\theta_c.
\end{equation}
That is somewhat lower than for positively charged particles [cf. Eq. (\ref{thetaVRinftypos111})], although the difference is not as significant as in the case of orientation (110) [cf. Eqs. (\ref{mean-thetaVR-harm-positive-110}) and (\ref{mean-thetaVR-harm-negative-110}) at $R/R_c\to\infty$].

%[which might as well be obtained from Eq. (\ref{thetaVRinfty-joint-parab-110}) by substituting $V_1\to V_L$, $V_2\to-V_S$], 

The slope of the $\sqrt{E}\overline{\chi}$ dependence on $E/R$ can readily be evaluated, too. In total, 
%\begin{equation}\label{F0-neg-111}
%V''(0)=-\frac{8V_L}{d_L^2}, \qquad F_0=\frac{4V_L}{d_L^2}d=5.7\frac{\text{GeV}}{\text{cm}},
%\end{equation}
%\begin{eqnarray}\label{kappa-neg-111}
%\kappa&=&\frac{4}{d_L}\sqrt{V_L}\int_{d_L/2}^{d/2}dr\ln\frac{d}{2r}\frac{d}{dr}\frac{r}{\sqrt{-\mathcal{V}_2(r) }}\nonumber\\
%&\approx& \frac{d_S}{d_L}\left(\frac{2V_S}{3V_L}+\frac{d_S}{d_L}\right)\approx 0.18.
%\end{eqnarray}
%In total,
\begin{eqnarray}\label{thetaVR-111-NLO}
\sqrt{\frac{E}{2}}\overline{\chi}
=\sqrt{V_L}\Bigg[1 -\frac{Ed_L}{4V_L R}\left(1+\ln\frac{4V_LR}{Ed}\right)\Bigg]\qquad\,\nonumber\\
+\frac{d_S}{\sqrt{V_S}}\left(\frac{V_L-V_S}{d}-\frac{E}{2R}\right)\text{arth}\sqrt{\frac{V_S}{V_L}}
+\mathcal{O}\left(\frac{R_c^2}{R^2}\right).
\end{eqnarray}

Eq. (\ref{thetaVR-111-NLO}) is valid for any values of $V_L$, $V_S$, $d_L$, $d_S$, provided the coefficient at $\text{arth}$ is positive: 
\begin{equation}\label{}
V_L>V_S+\frac{Ed}{2R}.
\end{equation}
Physically, that implies that the particle must have enough transverse energy to overpass the midway bump in the last interplanar interval of the bent crystal, in spite of the centrifugal tilt.
It is also straightforward to check that in the formal limit $d_S\to0$, the second line in (\ref{thetaVR-111-NLO}) vanishes, and the result boils down to Eq. (\ref{mean-thetaVR-harm-negative-110}) for a sequence of purely parabolic barriers [relevant for orientation (110)].
In the opposite formal limit $d_L\to0$ (a narrow major barrier), and $V_S\to0$ (a wide but shallow bump), (\ref{thetaVR-111-NLO}) goes over to the result (\ref{square-well-NLO}) for the square well to the NLO in $R_c/R$ [which was written there for positively charged particles in a high-$Z$ crystal in orientation (110)].

\subsubsection{Orientation (110), positively charged particles}\label{subsubsec:110poschar}

The pure parabolic model for orientation (110) considered in Sec. \ref{subsec:SimplestModels} does not take into account the smoothness of the continuous potential in vicinities of atomic planes (cf. Fig. \ref{fig:contin-pot}a below), which, as was shown in Sec. \ref{sec:gen-framework}, can give rise to a logarithmic modification of the linear $E/R$ dependence. This shortcoming can be amended by adopting a double parabolic potential, among the pieces of which, in contrast to the situation of Secs. \ref{subsubsec:111pos}, \ref{subsubsec:111neg}, one is convex upwards, while the other is convex downwards:
%Comparison of Eq.~(\ref{mean-thetaVR-harm-positive-110-NLO}) with experimental data for the mean VR angle in silicon in orientation (110), as conducted in \cite{Bond-VR-PRA}, had indicated that the predictions were in slight excess of the data at large $R$ (see Fig. \ref{fig:pos110}, dotted curve). 
%A self-suggesting improvement of this model is to let the potential to be piecewise parabolic within its period, accounting for smooth potential tops:
\begin{equation}\label{V-piecewise-parab}
V(r)=\begin{cases}
\mathcal{V}_1(r),  & 0\leq r\leq d_1/2,\\
\mathcal{V}_2(r),  & d_1/2\leq r\leq d-d_1/2,\\
\mathcal{V}_3(r),  & d-d_1/2\leq r\leq d.
\end{cases}
\end{equation}
Here 
\begin{equation}\label{cases-V2-def}
\mathcal{V}_2(r)=V_2\left(\frac{d-2r}{d_2}\right)^2
\end{equation}
is the harmonic bottom of the potential well, with $d_2<d$ and $V_2<V_0$, whereas parts
\begin{equation}\label{}
\mathcal{V}_1(r)=V_2+V_1-V_1\left(\frac{2r}{d_1}\right)^2,
\end{equation}
\begin{equation}\label{mathcalV3-def}
\mathcal{V}_3(r)=\mathcal{V}_1(d-r)=V_2+V_1-V_1\left(2\frac{d-r}{d_1}\right)^2,
\end{equation}
with $d_1=d-d_2\ll d$, $V_1=V_0-V_2\ll V_0$ describe a parabolic smearing of the potential tops in vicinities of points $r=0$ and $r=d$ correspondingly. Thus defined function $V(r)$ is continuous at $r=d_1/2$ and $r=d-d_1/2$. The requirement of its smoothness at those junctions imposes yet an additional condition $\mathcal{V}'_1(d_1/2)=\mathcal{V}'_2(d_1/2)$, equivalent to
\begin{equation}\label{junction-relation}
\frac{V_1}{d_1}=\frac{V_2}{d_2}=\frac{1}{4} F_{d_1/2}. 
\end{equation}
%The maximal value of the force, which is achieved in this model at the junctions $r=d_1/2$ and $r=d-d_1/2$, equals
%\begin{equation}\label{Fmax-double-parab}
%F_{d_1/2}=4\frac{V_0}{d}=4.43\frac{\text{GeV}}{\text{cm}},
%\end{equation}
%where we took the total well depth to be 
%\begin{equation}\label{V0-for-room-temp-Moliare}
%V_0=22.7\,eV 
%\end{equation}
%as for Moli\`{e}re's potential at room temperature, see Fig. \ref{fig:Moliere-vs-parab}a. The model value (\ref{Fmax-double-parab}) differs from the actual value $F_{\max}=6\frac{\text{GeV}}{\text{cm}}$ achieved for the Moli\`{e}re potential \cite{Bazylev-Glebov-Zhevago-book}, but that is not so important for VR. 
The adjustable parameters in this model are the total well depth $V_0$ and the shape parameter
\begin{equation}\label{delta-def}
\delta=\frac{d_1}{d}=\frac{V_1}{V_0}.
\end{equation}
%Its realistic value for the room temperature may be taken to equal $\delta=\frac{8V_0}{|V''(0)|d^2}\approx0.06$ (see Fig. \ref{fig:Moliere-vs-parab}, dashed curves).

%\begin{figure}
%\includegraphics{Moliere-vs-parab-letters-aa}
%\includegraphics{Moliere-vs-parab-letters-bb}
%\caption{\label{fig:Moliere-vs-parab} Interplanar continuous potential of a silicon crystal in orientation (110). (a) Solid curve, the sum of Moli\`{e}re potentials for individual atoms with a room-temperature coordinate distribution. Dashed -- piecewise parabolic potential (\ref{V-piecewise-parab}) having the same behavior near the tops, corresponding to $\delta=d_1/d=0.06$. 
%Dotted -- single parabolic, Eq. (\ref{V-pos-pure-parab}). 
%(b) The same for a smeared Moli\`{e}re potential, corresponding to $\delta=d_1/d=0.084$. }
%\end{figure}

Potential (\ref{V-piecewise-parab}) may be substituted to Eq. (\ref{general-mean-thetaVR}). As long as $V_1$ is small, in practice it will normally satisfy the inequality 
\begin{equation}\label{V1>maxV}
V_{\text{eff}1}(d_1/2)>\underset{r>d_1/2}\max V_{\text{eff}}(r),
\end{equation}
wherewith the reflection point will belong to the domain of $\mathcal{V}_2$ rather than $\mathcal{V}_1$ [see Eq. (\ref{V-piecewise-parab})]. Condition (\ref{V1>maxV}) can be solved for $R$ as 
\begin{equation}\label{R<}
R<\frac{E}{4V_1}\left(\sqrt{d}+\sqrt{d_2}\right)^2,
\end{equation}
where the right-hand side is much greater than 
\begin{equation}\label{}
R_c=\frac{E}{F_{d_1/2}},
\end{equation}
provided $d_1\ll d_2$. Since condition (\ref{R<}) is usually met in practice, strictly speaking, in this model $R$ can not be regarded as asymptotically large. Therefore, the NLO formula does not strictly apply here, and to be rigorous, one has to return to Eq. (\ref{general-mean-thetaVR}).

In Appendix \ref{app:B} it is shown that a satisfactory approximation for $\overline{\chi}$ under the present conditions is
\begin{equation}\label{meanVRangle-pos110-sqrtV2}
\overline{\chi}
\approx\frac{\pi}{2}\sqrt{\frac{2V_2}{E}}\left(1-\frac{R_c}{R}\right)^2.
\end{equation}

Compared to Eq. (\ref{mean-thetaVR-harm-positive-110}), the main correction here stems from the overall factor $\sqrt{V_2}\approx\sqrt{V_0}\left(1-{\delta}/{2}\right)<\sqrt{V_0}$. 
At that, 
\begin{equation}\label{}
K=\frac{\pi^2}{8}(1-\delta).
\end{equation}
%This small suppressing correction improves the agreement with the experimental data (see Sec. \ref{sec:phenomenology}). 
It also appears that in spite of the smearing of the potential around the planes, the dependence of (\ref{meanVRangle-pos110-sqrtV2}) on $E/R$ under condition (\ref{R<}) does not contain a logarithmic factor, because the particle merely does not enter the smeared region at all.

%, making it rather satisfactory in Fig. \ref{fig:pos110} (dashed curve). It may be surprising, though, that this agreement is better than that for the potentials themselves (see Fig. \ref{fig:Moliere-vs-parab}), so, the appropriate calculation for a realistic potential is desirable, anyway.

The model (\ref{V-piecewise-parab}) is already sufficiently detailed to be confronted with experimental data \cite{Scand-linear,Scandale-PRST-2008,Rossi-Scandale-2015} available for silicon crystal in this orientation. 
Its fit used in the NLO equation (\ref{NLOthetaVR-r*(ER)}) gives parameter values
\begin{equation}\label{fit:V0delta}
V_0=20.2\pm0.4 \text{ eV}, \quad \delta=0.042\pm0.026.
\end{equation}
It is also close, with nearly the same but less tightly constrained parameters, to the calculation by exact formula (\ref{general-mean-thetaVR}) for model (\ref{V-piecewise-parab}).

%The fitted model (shown in Fig. \ref{fig:pos110-fits} by the red dot-dashed curve) appears to be virtually indistinguishable from the model-independent linear regression (see Fig. \ref{fig:pos110-fits}, black dashed curve).  (Fig. \ref{fig:pos110-fits}, blue solid curve), which exhibits a visible curvature in the covered region. A slight difference occurs because the spanned region of $R$ here goes beyond $1/4R_c$, giving rise to next-to-next-to-leading order (NNLO) corrections.

%\begin{figure}
%\includegraphics{pos110-fits}
%\caption{\label{fig:pos110-fits} Crystal curvature dependence of the mean VR angle for positively charged particles in silicon in orientation (110) at energy $E=400$ GeV. Points: blue, \cite{Scand-linear}; green, \cite{Scandale-PRST-2008}; red, \cite{Rossi-Scandale-2015}. Curves: black dashed, linear regression; red dot-dashed, fit by double-parabolic model potential (\ref{V-piecewise-parab})--(\ref{mathcalV3-def}) used in the NLO equation (\ref{NLOthetaVR-r*(ER)}); blue solid, evaluation by exact Eq. (\ref{general-mean-thetaVR}) with potential (\ref{V-piecewise-parab})--(\ref{mathcalV3-def}). The $R_c=\frac{Ed}{4V_0}$ value was evaluated for the double-parabolic model potential.
%}
%\end{figure}

\subsubsection{Orientation (110), negatively charged particles}\label{subsubsec:110neg}

To obtain from (\ref{V-piecewise-parab}) the potential
for negatively charged particles, there is no need to flip it and shift by half the period -- it suffices just to replace $\delta$ by $1-\delta$, which interchanges the small and large parameters: $d_1\leftrightarrow d_2$, $V_1\leftrightarrow V_2$. 
Since a negatively charged particle traverses all the three regions of piecewise-parabolic potential (\ref{V-piecewise-parab}), the result of the exact calculation is more cumbersome. But the NLO calculation is simple. 
Evaluation by Eqs. (\ref{mean-thetaVR-infty}) and (\ref{V-piecewise-parab})--(\ref{delta-def}) gives
\begin{eqnarray}\label{chi0=thetac(1+delta2)}
\chi_0
&=&\sqrt{\frac{2V_2}{E}}\left[1+\frac{d_1}{d}\left(\sqrt{\frac{V_1}{V_2}}+\sqrt{\frac{V_2}{V_1}}\right)\text{arth}\sqrt{\frac{V_1}{V_2}}\right]\nonumber\\
&\simeq& %\sqrt{\frac{2V_2}{E}}\left(1+\frac{d_1}{d}\right)=
\theta_c\left(1+\delta/2\right),\qquad\qquad\quad
\end{eqnarray}
with $\theta_c=\sqrt{2(V_1+V_2)/E}$ and $\delta=d_1/d\ll 1$, while evaluation of the slope can be done straightforwardly by Eq. (\ref{NLOthetaVR-r*(ER)}):
\begin{equation}
\overline{\chi}=\chi_0\left\{1-\frac{R_c}{R}\left[1+(1-\delta)\ln\frac{R}{R_c}\right]\right\}.
\end{equation}
That explicates the $\delta$-dependent correction to Eq. (\ref{mean-thetaVR-harm-negative-110}). 
Naturally, the increase of $\delta$ in turn increases $\chi_0$ [in contrast to the case of positively charged particles -- cf. Eq. (\ref{meanVRangle-pos110-sqrtV2})], whereas the slope diminishes, because the potential well broadens [see the commentary after Eq. (\ref{mean-thetaVR-NLO-sharpedge-ab})].

The expressions for $\overline{\chi}$ obtained in this subsection for double-parabolic model potentials exhibit surprisingly simple dependencies on various typical parameters. They can even provide rather accurate phenomenological predictions, which can be used for guidance in practice. But the meaning of such coincidences should not be overestimated.

%\newpage

%.

%

\section{$R\to R_c$ limit}\label{sec:Rtorc}

It may also be of interest to determine the behaviour of $\overline{\chi}$ near the critical point $R=R_c$, although it can hardly be of high practical value, insofar as $\overline{\chi}$ there tends to zero. 
In the previous section we had seen that for parabolic potential models, $\overline{\chi}$ vanishes at the threshold quadratically: 
\begin{equation}\label{chi-sim-(1-R_c/R)^2}
{\overline{\chi}}/{\theta_c}\underset{R\to R_c}\sim (1-R_c/R)^2.
\end{equation}
But those models presume that the second derivative of the effective potential remains constant over the entire interplanar interval. In reality, the potential is smooth near the atomic planes. 
At $R\to R_c$ the potential well is tilted so strongly that the radicand of (\ref{general-mean-thetaVR}) is positive only in a small vicinity of point $r_s$, wherein $V'(r_s)$ is maximal, i.e., $V''(r_s)=0$. 
To describe the effective potential under such a condition, one thus has to expand $V(r)$ up to the third order in $r-r_s$:
\[
V_{\text{eff}}(r)=V_{\text{eff}}(r_s)+(r-r_s)\!\left(\! V'(r_s)-\frac{E}{R}\right)+\frac16 (r-r_s)^3 V'''(r_s).
\]
Here $V'(r_s)=F_{\max}>0$, and $V'''(r_s)<0$. 
Therewith, integral (\ref{general-mean-thetaVR}) assumes the form\footnote{We will be specifically talking here about the single-well case (110). For the double-well case (111), critical points in both wells are reached almost simultaneously, so, their contributions should be added.}
\begin{equation}\label{}
\frac{d}{2}\overline{\chi}(E,R) 
=\sqrt{\frac{|V'''(r_s)|}{3E}}\!\int_{-2\sqrt{\lambda/3}}^{\sqrt{\lambda/3}}dr'
\sqrt{ 2\left({\lambda}/{3}\right)^{3/2}\!-\!\lambda r'\!+\!r'^3}
\end{equation}
with $r'=r-r_s$ and $\lambda=\frac{6F_{\max}}{|V'''(r_s)|}\left(1-\frac{R_c}{R}\right)$.
Evaluating it, we get
\begin{eqnarray}\label{chi-sim-(1-RRc)54}
\overline{\chi}\underset{R\to R_c}\simeq\frac{24}{5d}2^{5/4} \left(\frac{E}{|V'''(r_s)|}\right)^{3/4} 
\left(\frac{1}{R_c}-\frac{1}{R}\right)^{5/4}.
\end{eqnarray}
This relation is valid both for positively and for negatively charged particles. 

Compared to (\ref{chi-sim-(1-R_c/R)^2}), the decrease law in (\ref{chi-sim-(1-RRc)54}) is steeper, but the prefactor involves $\left|V'''(r_s)\right|^{-3/4}\sim u^{3/2}$, with $u$ being the rms thermal displacement of atoms off the plane [see Eqs. (\ref{VP=conv-Gauss-VPS}) and (\ref{u075}) below], whereby ratio 
\begin{equation}\label{}
\frac{\overline{\chi}}{\theta_c}\underset{R\to R_c}\sim\left(\frac{u}{d}\right)^{3/2} \left(1-\frac{R_c}{R}\right)^{5/4}
\end{equation}
is suppressed by a small factor $(u/d)^{3/2}$. Therefore, asymptotic law (\ref{chi-sim-(1-RRc)54}) holds only in a narrow vicinity of $R_c$, whereas at $R\sim 4R_c$ the behaviour of ${\overline{\chi}}/{\theta_c}$ may be more similar to (\ref{chi-sim-(1-R_c/R)^2}).

%\newpage

\section{Use of realistic potentials}\label{sec:phenomenology}

Let us finally turn to a more realistic description of the mean VR angle, 
based on exact generic representations (\ref{general-mean-thetaVR}), (\ref{general-mean-thetaVR-two-wells}) for $\overline{\chi}$. 
%It will also be instructive to check how accurate are the NLO formulae of Sec. \ref{sec:expansion-RcR}.
Our knowledge of the interplanar potential, 
even though good, is in principle only approximate, and its uncertainty gives rise to some uncertainties in theoretical predictions. 
We will complete our study by assessing this sensitivity. 
That can be done best in comparison of predictions obtained with different potentials with experimental data. 
A comparison of experimental data with the potential deflection theory was formerly presented in \cite{,Bellucci-Chesnokov-Maisheev-Yazynin} for positively charged particles. 
Later, there were also published detailed data for negatively charged particles reflecting on a crystal in orientation (111) \cite{Wistisen-Wienands}. 
Along with the phenomenological check of exact equations, it is important yet to assess the accuracy of NLO approximations derived in Sec. \ref{sec:expansion-RcR}.

%expressions for $\overline{\chi}$ derived in Sec. \ref{sec:gen-framework}, besides facilitating the calculations, can be employed for a more fundamental purpose of directly relating the experimental data to parametrizations for the interplanar potentials, and thereby refining their parameters. Although atomic potentials have long been the subject of phenomenological studies and ab initio calculations, there thus far seems to be no commonly accepted, ``standard'' parameterization for the potential serving perfectly in all physical situations. 

\subsection{World data. Intercepts and slopes}

Experimental data on VR, obtained for various crystals and their orientations, for various incident particles and their energies, can be combined together by plotting $\sqrt{E}\overline{\chi}$ vs. $E/R$. 
Up-to-date data (see Figs. \ref{fig:pos110-Moliere-DT}--\ref{fig:neg111-Moliere-DT} below) do not exhibit a marked curvature, basically being compatible with linear dependencies. 
Thus, for those cases the logarithmic modification of the slope calculated in Sec. \ref{sec:expansion-RcR} and present in theoretical predictions for negatively charged particles in Figs. \ref{fig:neg111-Moliere-DT}, \ref{fig:neg110-Moliere-DT} is not revealed yet, because of the scatter between the points. 
Nonetheless, two numbers inferred from experimental data (the intercept and the slope of the linear dependence) for each particle charge sign and crystal orientation together can appreciably constrain the interplanar potential.

The tightest experimental constraint is for $\chi_0$, it corresponds to an accuracy of a few percent (standard deviation) -- see Table \ref{table:thetainf} below. The slope $\sqrt{E}\frac{d(\sqrt{E}\overline{\chi})}{d(E/R)}$ is constrained more loosely -- only to within 5--30\%. 
Nonetheless, it can already rule out the crudest models, such as the pure parabolic potential for (110) orientation, containing only one parameter. For instance, for 400 GeV protons on Si (110) \cite{Scand-linear,Scandale-PRST-2008,Rossi-Scandale-2015} from the linear regression to the experimental data one infers
\begin{equation}\label{089}
%\frac{\chi_0}{2d}\left|\frac{\partial\overline{\chi}}{\partial(1/R)}\right|_{E/R\to0}
K=0.89\pm0.05,
\end{equation}
which is more than 6 standard deviations below the corresponding pure parabolic potential model value (\ref{pi28}). 
It must be realized, though, that the experimentally covered range of curvatures extends somewhat beyond $1/4R_c$, wherewith the NLO approximation itself becomes insufficiently accurate, and the value of $K$ determined in such a broad interval of $1/R$ appears to be underestimated due to sizable NNLO corrections. 
The use of Eq. (\ref{mean-thetaVR-harm-positive-110}) valid beyond the NLO improves the agreement, but does not make it perfect, anyway. 
%With the increase of accuracy of experimental data, the pure parabolic potential model is expected to be eventually excluded.
A double parabolic potential model of Sec. \ref{subsubsec:110poschar} has two adjustable parameters, by virtue of which it can sustain statistical comparison with VR experiments better.

%Given the modest precision of the world experimental data, one may anticipate that they can hardly compete with the accuracy of theoretical predictions, and will be reproducible basically by any realistic potential. Indeed, the mean VR angle is determined by an integral (\ref{general-mean-thetaVR-max}), and thus is likely to be sensitive only to gross features of the integrand. 
%Nonetheless, let us check whether that is really the case. 

\subsection{Parametrizations for atomic potentials}

Most valuable, of course, would be to compare the experimental data with predictions based on the best available potential.
There are ab initio calculations thereof at a varying degree of sophistication, but the crystal structure is in principle more complicated than that of its constituent atoms. 
Therefore, a commonly adopted  attitude for construction of crystal potentials is just to add potentials for isolated atoms, 
\begin{equation}\label{V-g-def}
V_{1a}(r)=\frac{Z_1Z e^2}{r}g(r), \quad g(0)=1, \quad g(r)\underset{r\to\infty}\to0,
\end{equation}
with $Z_1$ being the projectile charge in units of proton charge $e$, by
placing them at the lattice sites, and neglecting interaction between the atoms, i.e., solid-state binding effects. 
When the number of covalently paired electrons is much smaller than the total electron number, their effect is expected to be relatively small. 
For silicon,  there are 4 valence vs. total 14 electrons (although half of the covalent bonds is extended along the considered low-index planes), so, the validity of this approximation may be not self-evident . Nonetheless, in this paper we confine our objective to comparing predictions from noninteracting-atom potentials.

The potential of a thus obtained atomic plane has the form
\begin{eqnarray}\label{VPS-intV}
V_{PS}(x)&=&n_s\int d^2r_{\perp} V_{1a}\left(\sqrt{x^2+r_{\perp}^2}\right)\nonumber\\
&=&2\pi n_s\int_{|x|}^{\infty}drr V_{1a}\left(r\right),
\end{eqnarray}
where $n_s$ is the areal density of atoms in the plane.
It is also necessary to average it over the thermal and zero-point fluctuations of the atom positions, 
convolving with a Gaussian distribution \cite{Appleton,Gemmell}:
\begin{equation}\label{VP=conv-Gauss-VPS}
V_P(x)=\int_{-\infty}^{\infty}dx'\frac1{\sqrt{2\pi}u}e^{-x'^2/2u^2}V_{PS}(x-x'),
\end{equation}
where the rms $x$-projection of atom displacement at room temperature can be inferred directly from the dedicated experiments \cite{u075}:
\begin{equation}\label{u075}
    u=0.075 \text{ \AA} \qquad (\text{silicon,\quad} T=300\text{ K}).
\end{equation}
Finally, the potential of the entire crystal for orientation (110) is built from (\ref{VPS-intV}), (\ref{VP=conv-Gauss-VPS}) as\footnote{Albeit at evaluation of the VR angle the integration in (\ref{theta-VR-zeta}), (\ref{general-mean-thetaVR-max}) and in the subsequent equations is carried out over a single period of the crystal, in (\ref{Vcryst=sum}) it is necessary to take a sufficiently large number of planes in order to ensure the symmetry of the potential wrt any of its minima and maxima.}
\begin{equation}\label{Vcryst=sum}
V(x)=\sum_{n=-\infty}^{\infty}V_P(x+nd),
\end{equation}
where $d=1.92$ \AA \, is the interplanar distance, whereas for orientation (111),
\begin{equation}\label{V111=sum}
V(x)=\sum_{n=-\infty}^{\infty}V_P\left\{x+d_S\left(n+2\lfloor n/2\rfloor\right)\right\},
\end{equation}
where $d_S=0.78\text{ \AA}$ is the shorter interplanar distance (cf. Secs. \ref{subsubsec:111pos}, \ref{subsubsec:111neg}), and $\lfloor n/2\rfloor$ is the (lower) integer part of $n/2$. 

One of the most commonly used parametrizations for the atomic potential screening function is that proposed long ago by Moli\`{e}re \cite{Moliere} as an analytic interpolation to the Thomas-Fermi approximation:
\begin{equation}\label{Moliere-3param}
g(r)\approx\sum_{k=1}^3 \alpha_k e^{-\beta_k r/a_{TF}}, \qquad \alpha_k, \beta_k>0,\qquad \sum_{k=1}^3\alpha_k=1.
\end{equation}
Accordingly, it depends on the ratio $r/a_{TF}$, where the Thomas-Fermi radius $a_{TF}(Z)$ equals
\begin{equation}\label{aTF-0}
    a_{TF}=0.885 a_{B} Z^{-1/3}=0.194\text{ \AA} \quad \text{for}\quad Z=14,
\end{equation}
with $a_B$ the Bohr radius. 
But it is not devoid of shortcomings: the Poisson equation
for a spherically-symmetrical electron density $n_e(r)$,
\begin{equation}\label{g''proptor}
g''(r)=\frac{4\pi }{Z} r n_e(r),\qquad \int d^3 n_e(r)=Z,
\end{equation}
implies that $g''(r)$ along with the product $r n_e(r)$, where $n_e(r)$ is everywhere finite, 
must vanish\footnote{For the static continuous potential of a single plane, according to Eq. (\ref{VPS-intV}), that implies $V'''_{PS}(0)=0$.} at $r=0$:
\begin{equation}\label{g''(0)=0}
g''(0)=0.
\end{equation}
In contrast, Moli\`{e}re parametrization gives $g''(0)=a_{TF}^{-2}\sum \alpha_k \beta_k^2 >0$, since all $\alpha_k>0$. 

\begin{figure}
\includegraphics[width=0.86\columnwidth]{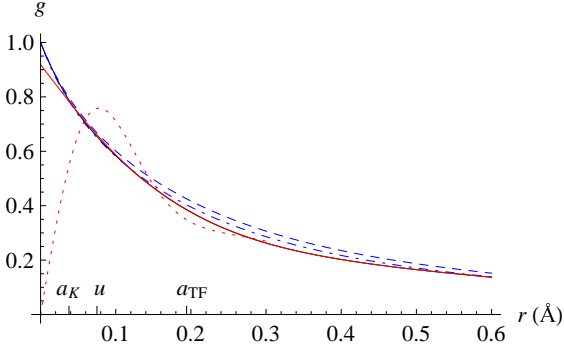}
\caption{\label{fig:g} Screening function parametrizations for an isolated silicon atom. 
Black solid curve, Lobato-Van Dyck potential (\ref{Lobato-VanDyck-g}). 
Red solid, $f_X$-based Doyle-Turner potential (\ref{Doyle-Turner-fX}). 
Red dotted, $f_e$-based Doyle-Turner potential (\ref{DT-fe-Gaussparam}). 
Black dashed, Moli\`{e}re potential (\ref{Moliere-3param}). 
Black dot-dashed, Moli\`{e}re potential (\ref{Moliere-3param}) with modified screening radius (\ref{aTF-1}).
The axis ticks show characteristic atomic scales, such as the K-shell radius $a_K=a_B/Z$ and Thomas-Fermi radius (\ref{aTF-0}), as well as the rms atom thermal displacement in a solid state (\ref{u075}).
}
\end{figure}

There is experimental evidence (see, e.g., \cite{Schwabe-Stolle}) that Moli\`{e}re parametrization predicts an excessive screening radius. There were proposed various heuristic recipes for reducing it \cite{Gemmell,SRIM-book}, for instance 
\cite{Bohr1948,LNS},
\begin{equation}\label{aTF-1}
    \tilde a_{TF}=0.885 a_{B} \left(1+Z^{2/3}\right)^{-1/2}=0.179\text{ \AA} \quad \text{for}\quad Z=14.
\end{equation}
%But it does not cure the mentioned intrinsic flaw of the parametrization itself.

The most accurate approach for calculation of atomic structure is relativistic Hartree-Fock (HF). 
Two commonly used parameterizations of HF calculations of atomic structure were proposed by Doyle and Turner \cite{Doyle-Turner}.\footnote{In this, relatively old version of HF calculation, electron correlation effects were neglected. More modern, multi-configuration HF techniques \cite{Rez,Meyer-Muller-Schweig} had assessed those effects on the mean electron density to be $\sim0.1\%$.}
One of them is for the electron scattering form factor
\begin{equation} 
f_e(s)=\sum_{k=1}^4 a_k e^{-b_k s^2}, \qquad s=\frac{\sin\theta}{\lambda},
\end{equation} 
with $a_k$, $b_k$ given in Table 4 of \cite{Doyle-Turner},
which is convenient for deriving the mean electrostatic potential: 
\begin{eqnarray}\label{DT-fe-Gaussparam}
\frac{Z g(r)}{r}&=&\frac{16\pi a_B}{r} \int_0^{\infty}dss f_e(s) \sin 4\pi sr\nonumber\\
&\approx& 16\pi^{5/2} a_B\sum_{k=1}^4 \frac{a_k}{b_k^{3/2}}e^{-\frac{4\pi^2 r^2}{b_k}}.
\end{eqnarray}
It satisfies condition (\ref{g''(0)=0}), but unfortunately, is plagued by a more severe intrinsic flaw: $g(0)=0$ instead of 1 (see Fig. \ref{fig:g}, dotted curve), and $g'''(0)=0$ instead of value $\frac{4\pi}{Z}n_e(0)$ following from Poisson equation (\ref{g''proptor}).
This parametrization is nonetheless often used for channeling-related problems, including VR \cite{Wistisen-Wienands,Sytov-SubGev,Scand-Tar-PhysRep}.

Another Doyle-Turner parametrization is 
for X-ray scattering factor: 
\begin{eqnarray}\label{Doyle-Turner-fX}
    f_X(s)&=& \frac{1}{s}\int_0^{\infty} dr r n_e(r) \sin 4\pi s r \nonumber\\
    &\approx&\sum_{k=1}^4 a_k e^{-b_k s^2}+c,
\end{eqnarray}
where $a_k$, $b_k$, and $c=Z-\sum_{k=1}^4 a_k>0$ are given in Table 3 of \cite{Doyle-Turner} or Table 6.1.1.4 of \cite{Intern-Tables-Cryst}.
Constant $c\sim 1$ is arguably non-physical, since it contradicts the correct asymptotic behaviour $f_X(s)\underset{s\to\infty}\simeq \frac{|n'_e(0)|}{2^5 \pi^3 s^4}$, but parametrization (\ref{Doyle-Turner-fX}) is intended to hold only for $s\leq 2\text{ \AA}^{-1}$. 
%For greater $s$, it requires some extrapolation, which, however, can not be done quite self-consistently by Gaussians.
In its original form, extended to arbitrary $s$,
\begin{eqnarray}\label{g-fX}
g(r)&=&1-\frac{2}{\pi Z}\int_0^{\infty}\frac{ds}{s}f_X(s)\sin 4\pi sr\nonumber\\
&\approx&\frac{1}{Z}\sum_k a_k \text{erfc}\left(\frac{2\pi r}{\sqrt{b_k}}\right),
\end{eqnarray}
where $\text{erfc}$ is the complementary error function, it was used for description of VR in \cite{Maisheev,Bellucci-Chesnokov-Maisheev-Yazynin}.
It obeys requirement (\ref{g''(0)=0}), but predicts $g(0)=1-c/Z<1$ (i.e., $c$ out of $Z$ electrons sitting exactly at the nucleus),\footnote{Alternatively, one can let $f_X(s)=0$ for $s> 2\text{ \AA}^{-1}$. That fulfils condition $g(0)=1$, but gives rise to an oscillatory component in $g(r)$ at $r>0$.}
 and vanishing initial fourth derivative: $g^{(\text{iv})}(0)=0$, instead of $g^{(\text{iv})}(0)=\frac{4\pi}{Z}n'_e(0)$ following from Poisson equation (\ref{g''proptor}).

\begin{figure}
\includegraphics[width=0.98\columnwidth]{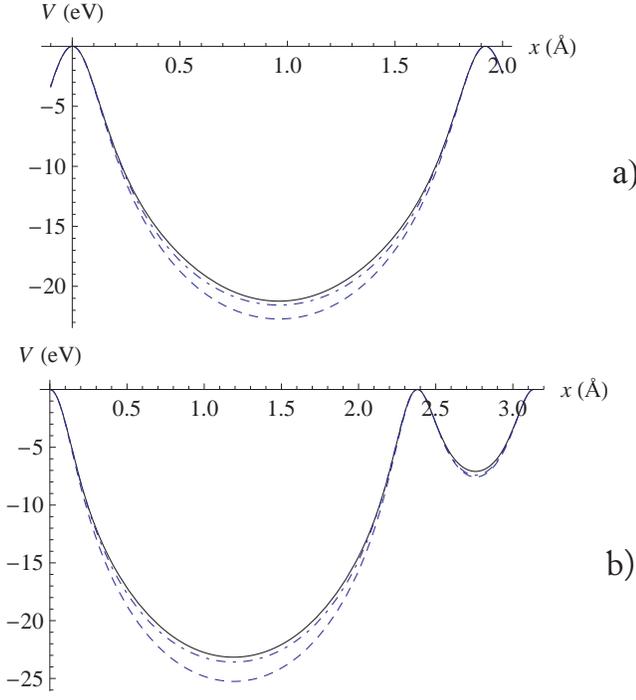}
\caption{\label{fig:contin-pot} Planar continuous potentials over one period for positively charged particles in a silicon crystal at room temperature, evaluated via Eqs. (\ref{V-g-def})--(\ref{V111=sum}): a). in orientation (110); b). in orientation (111). Black solid curves, HF calculations [by any of Eqs. (\ref{DT-fe-Gaussparam}), (\ref{Doyle-Turner-fX}) or (\ref{Lobato-VanDyck-g})]. Blue dashed curves, Moli\`{e}re potential, Eqs. (\ref{Moliere-3param}), (\ref{aTF-0}). Blue dot-dashed, Moli\`{e}re potential with modified screening radius (\ref{aTF-1}).
}
\end{figure}

A completely self-consistent parametrization of HF data,
\begin{equation} 
n_e(r)\approx 2\pi^4 a_B\sum_{k=1}^5\frac{a_k}{b_k^{5/2}}e^{-\frac{2\pi r}{b_k^{1/2}}}
\end{equation} 
was proposed in \cite{Lobato-VanDyck}. It consists of a sum of exponentials, like (\ref{Moliere-3param}) (permitting some of parameters $a_k$ to be negative), but is formulated for the electron density rather than the screening function. 
The latter derives as
\begin{equation}\label{Lobato-VanDyck-g}
g(r)\approx\frac{1}{Z}\sum_{k=1}^5 N_k \left(1+\frac{\mu_k r}{2}\right)e^{-\mu_k r},
\end{equation} 
with
\begin{equation}
N_k= 2\pi^2 a_B \frac{a_k}{b_k}, \quad \mu_k=\frac{2\pi}{\sqrt{b_k}},
\end{equation}
and the thermal-averaged potential of a single atomic plane reads
\begin{eqnarray}\label{LVD-1plane-T}
    V_P (x)\approx\frac{\pi n_s e^2}{2}\sum_{k=1}^5 N_k\Bigg\{2u\sqrt{\frac{2}{\pi}}e^{-x^2/2u^2}\qquad\qquad\qquad\quad\nonumber\\
+e^{\mu_k^2 u^2/2+\mu_k x}\left(\frac{3}{\mu_k}-\mu_k u^2-x\right)\text{erfc}\left[\frac{1}{\sqrt{2}}\left(\mu_k u+\frac{x}{u}\right)\right]\quad\nonumber\\
+e^{\mu_k^2 u^2/2-\mu_k x}\left(\frac{3}{\mu_k}-\mu_k u^2+x\right)\text{erfc}\left[\frac{1}{\sqrt{2}}\left(\mu_k u-\frac{x}{u}\right)\right]\!\Bigg\}.\nonumber\\
\end{eqnarray} 

Although screening functions of HF potentials described above are markedly different at small $r$ (see Fig. \ref{fig:g}), the corresponding continuous planar potentials obtained after integration over longitudinal coordinates (\ref{VPS-intV}) and thermal averaging (\ref{VP=conv-Gauss-VPS}) are visually indistinguishable.\footnote{The amplitude Debye-Waller factor $e^{-(4\pi s u)^2/2}$ with $u$ given by Eq. (\ref{u075}) and $s=2\text{ \AA}^{-1}$ equals 0.17, which is small enough to completely suppress already small values of the form factor in that region.} 
In Figs. \ref{fig:contin-pot}a,b they correspond to the same solid curves. 
The Moli\`{e}re continuous potential, though, deviates from HF markedly.
The reason is that Moli\`{e}re screening function deviates from HF at large $r$, whereas small-$r$ variations are less important, being smeared out by averaging procedures, anyway. 
However, Moli\`{e}re potential with modified screening radius (\ref{aTF-1}), even though fortuitously, is very close to HF, as well. 
In this sense, it can be regarded as competitive.

%The temperature dependence enters there via the same Gaussian convolution. 

%Doyle-Turner potential is numerically fairly close to Moli\`{e}re's one, and is also used in simulations of channeling and VR .

%\newpage

\subsection{Comparison with experiments}\label{subsec:experim}

\begin{figure}
\includegraphics{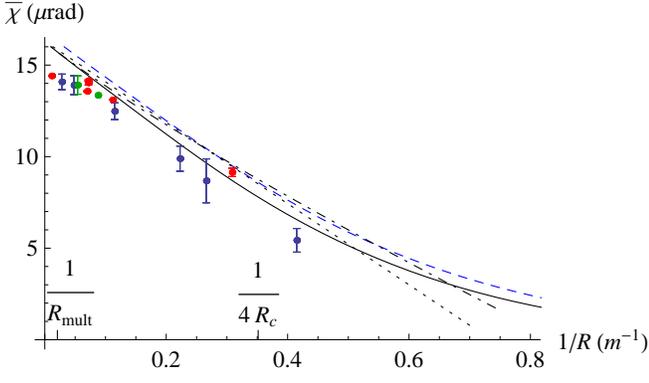}
\caption{\label{fig:pos110-Moliere-DT} Dependence of the mean VR angle for positively charged particles with energy $E=400$ GeV in silicon crystal in orientation (110) on the crystal curvature. 
Points, measurements in different experiments: blue, \cite{Scand-linear}; green, \cite{Scandale-PRST-2008}; red, \cite{Rossi-Scandale-2015}. Curves: blue dashed, Moli\`{e}re potential [with ordinary screening radius (\ref{aTF-0})] at room temperature [Eq. (\ref{u075})]; black solid, HF potential (\ref{LVD-1plane-T}); black dot-dashed, NLO prediction for HF potential, Eqs. (\ref{NLOthetaVR-r*(ER)}), (\ref{r*-def}). 
Black dotted, static NLO prediction for HF potential, Eq. (\ref{NLO-Vstatic}).
}
\end{figure}

\begin{figure}
\includegraphics{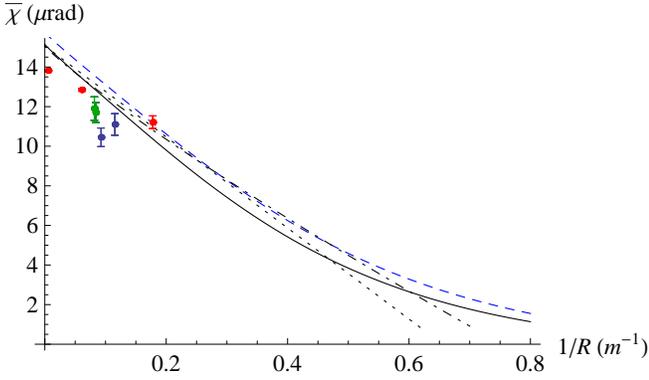}
\caption{\label{fig:pos111-Moliere-DT} CDependence of the mean VR angle for positively charged particles with energy $E=400$ GeV in silicon crystal in orientation (111) on the crystal curvature. 
Points: blue, \cite{Scandale-PRST-2008}; green, \cite{ScandalePLB111}; red, \cite{Rossi-Scandale-2015}. Curves: the same as in Fig. \ref{fig:pos110-Moliere-DT}.
}
\end{figure}

\begin{figure}
\includegraphics{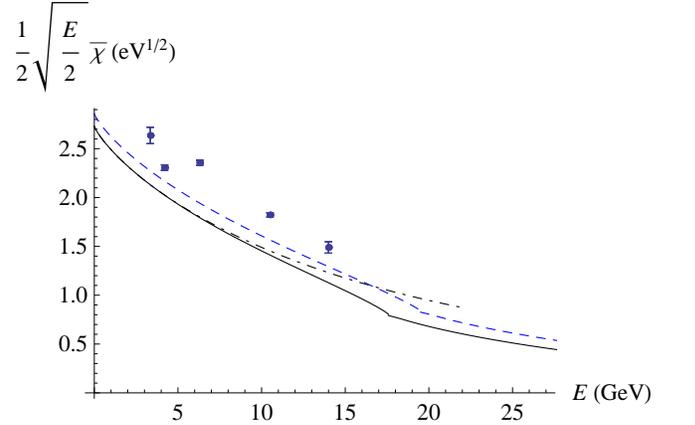}
\caption{\label{fig:neg111-Moliere-DT} Energy dependence of the mean VR angle for negatively charged particles in silicon crystal in orientation (111) at $R=15$ cm. Points, experimental data \cite{Wistisen-Wienands}. Curves: the same as in Fig. \ref{fig:pos110-Moliere-DT}. 
A break in the curves around $E=18$ GeV occurs because the minor potential bump rises above the major barrier due to the large tilt, making the potential double-well instead of single-well. At that, one has to switch from Eq. (\ref{general-mean-thetaVR}) to Eq. (\ref{general-mean-thetaVR-two-wells}).
}
\end{figure}

\begin{figure}
\includegraphics{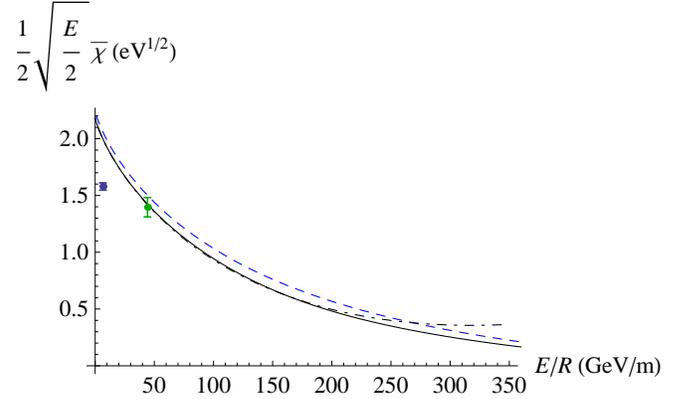}
\caption{\label{fig:neg110-Moliere-DT} $E/R$-dependence of the mean VR angle for negatively charged particles in silicon crystal in orientation (110). Points: blue, measurement \cite{Scandale-neg} for $E=150$ GeV and $R=22.79$ m; green, measurement \cite{Bandiera-VR-rad} for $E=120$ GeV and $R=2.71$ m.
Curves: the same as in Fig. \ref{fig:pos110-Moliere-DT}. 
}
\end{figure}

The theoretical predictions for the mean VR angle computed by Eq. (\ref{general-mean-thetaVR}) or (\ref{general-mean-thetaVR-two-wells}) with Moli\`{e}re and HF potentials, as well as NLO predictions for HF  potential computed by Eqs. (\ref{NLOthetaVR-r*(ER)}), (\ref{r*-def}), are compared with the experimental data in Figs. \ref{fig:pos110-Moliere-DT}--\ref{fig:neg110-Moliere-DT}. 
Given that room-temperature planar potentials for all types of HF screening functions are very close (Fig. \ref{fig:contin-pot}), it is natural that their predictions for the mean VR angle are indistinguishable in the figures.
It can be seen from the plots that the HF potential predictions are generally the most accurate, whereas the original Moli\`{e}re potential, as was previously pointed out in \cite{Bellucci-Chesnokov-Maisheev-Yazynin}, tends to give excessive  predictions. 
But its relative deviation is twice smaller than that for the potential well depths in Fig. \ref{fig:contin-pot}, which is natural in view of the square root dependence on the potential in Eqs. (\ref{general-mean-thetaVR}), (\ref{general-mean-thetaVR-two-wells}). 
So, the sensitivity of $\overline{\chi}$ to the interplanar potential shape can be regarded as moderate. 
The predictions of the Moli\`{e}re potential with modified screening radius (\ref{aTF-1}), in fact, virtually fall on top of the HF curves, too.

The curve slopes depend on the atomic potential weakly, so, the main differences in their values stem from the differences in $\chi_0$. The latter, according to Eqs. (\ref{2localthetac})--(\ref{thetac(r)-def}), boils down to $\left\langle\sqrt{\max V-V(r)}\right\rangle_r$.  
The values of $\left\langle\sqrt{\max V-V(r)}\right\rangle_r$ inferred from a linear fit to the experimental data  for $\frac12\sqrt{\frac{E}{2}}\chi_0$, as well as from theoretical predictions, are listed in Table \ref{table:thetainf}. 
This table allows one to assess subtle differences between the predictions of different potentials.



% for  calculated by Eqs. (\ref{2localthetac})--(\ref{thetac(r)-def}) with the experimental data being inferred from the  (see ),\footnote{The data for VR of negatively charged particle in (110) crystal orientation are too scarce to permit reliable statistical analysis. The trend for them is qualitatively the same as for negatively charged particles and orientation (110).} shows that predictions of the Doyle-Turner potential for positively charged particles exceed the experimental data by more than 6 standard deviations. Its small deviation from the Moli\`{e}re's potential results in visible differences for the VR angle, but does not improve the agreement with the experiments. The Moli\`{e}re potential (see Table \ref{table:thetainf}) offers an even better agreement for $\chi_0$ of negatively charged particles and orientation (111), but similarly gives excessive predictions for positively charged particles.
%For negatively charged particles and orientation (110) (see Fig. \ref{fig:neg110-Moliere-DT}), the data are not enough to discriminate between the theoretical predictions.

It should be cautioned that the pure continuous potential deflection theory is valid under condition 
\begin{equation}\label{RllRmult}
R\ll R_{\text{mult}}(E),
\end{equation} 
where the parameter in the rhs, quantifying the onset of incoherent multiple scattering effects,\footnote{I.e., breakdown of condition 
\begin{equation}\label{rnmaxllsigmaam}
\sqrt{\Delta E_{\perp}/E}\gg \sigma_{\text{am}}(\Delta z),
\end{equation} 
where the lhs is the typical angle of atomic plane crossing by a VR particle near the radial reflection point, and the rhs is the rms incoherent multiple scattering angle acquired in the intrinsic VR region of the length $\Delta z$ defined by Eq. (\ref{Delta-z-VR}). Violation of condition (\ref{rnmaxllsigmaam}) will lead to impossibility for the particle to pass the entire sequence of potential wells, and hence to a decrease of the reflection angle.} 
may be evaluated by the engineering formula (64) of \cite{Bond-PRST}:
\begin{equation}\label{Rmult-def}
    R_{\text{mult}}=\left(\frac{E}{18\text{ GeV}}\right)^{5/4}[\text{m}].
\end{equation} 
At $R^{-1}\lesssim R_{\text{mult}}^{-1}$, the experimental points in Figs. \ref{fig:pos110-Moliere-DT}--\ref{fig:pos111-Moliere-DT}  subside below the theoretical prediction. Quantitative explanation of this effect is challenging for the analytic theory.

The logarithmically modified NLO approximation developed in Sec. \ref{subsec:NLO-log-modif} (blue dot-dashed curves in Figs. \ref{fig:pos110-Moliere-DT}--\ref{fig:neg110-Moliere-DT}) proves to work nicely for negatively charged particles (Figs. \ref{fig:neg111-Moliere-DT}--\ref{fig:neg110-Moliere-DT}). 
For positively charged particles, its application leads to somewhat worse, slightly \emph{excessive} predictions (compared with the exact result) in the intermediate region. 
This resembles the situation with a square-well, rather than parabolic potential (see the end of Sec. \ref{subsec:SimplestModels}), although visually interplanar potentials in Fig. \ref{fig:contin-pot} look close to parabolic. 
That occurs because, as was mentioned in Sec. \ref{subsec:NLO-log-modif}, the thermally smeared potential around the atomic planes is not described by a reverted parabola at a long extent, and at intermediate $R_c/R$, radial reflection points belong to a non-parabolic region. 
%Nonetheless, predictions of the NLO approximation are not worse than those with the exact formula (\ref{general-mean-thetaVR-max}) and Moli\`{e}re potential.

If instead one employs the NLO formula for the static potential [Eq. (\ref{NLO-Vstatic})], which displayed in Figs. \ref{fig:pos110-Moliere-DT}, \ref{fig:pos111-Moliere-DT} by dotted curves, it proves to be closer to the exact result at moderate $R_c/R$, while being somewhat worse at $R_c/R\to 0$, which is not surprising since this approximation is not truly asymptotic. 
But physically, at very small $R_c/R$, as was mentioned above, the pure potential description  of deflection is invalid, anyway.
Therefore, in practice, the use of the static-potential NLO approximation may have certain advantages.

\begin{center}
\begin{table}[h]
\caption{\label{table:thetainf} Mean values of $\left\langle\sqrt{\max V-V(r)}\right\rangle_r$ (in eV$^{1/2}$) as determined from experimental  data on $\frac12\sqrt{\frac{E}{2}}\chi_0$ 
\cite{Scand-linear,Scandale-PRST-2008,Rossi-Scandale-2015,Wistisen-Wienands,Scandale-PRST-2008,ScandalePLB111}, in comparison with predictions based on model potentials. Moli\`{e}re potential corresponds to the screening radius evaluated by Eq.~(\ref{aTF-0}). Moli\`{e}re with modified $a_{TF}$ is evaluated via Eq. (\ref{aTF-1}).}
%\footnotesize\rm
\centering
%\begin{ruledtabular}
\begin{tabular}{|c|c|c|c|c|}%was {@{}*{8}{l}}
\hline
 & (110) pos & (111) pos & (111) neg \\
\hline
Experiments &$3.39\pm0.04$& $3.0\pm0.1$ &$2.9\pm0.1$\\
Moli\`{e}re&3.77&3.52&2.86\\
Moli\`{e}re, modified $a_{TF}$ & 3.68 & 3.43 & 2.74\\
Doyle-Turner, $f_e$-based &3.6437 & 3.387 & 2.734\\
Doyle-Turner, $f_X$-based &3.6439 & 3.389 & 2.729 \\
Lobato-Van Dyck & 3.6445 & 3.389 & 2.729\\
\hline
\end{tabular}
%\end{ruledtabular}
\end{table}
\end{center}

Finally, it attracts attention that experimental data for negatively charged particles, at least for orientation (111) (see Fig. \ref{fig:neg111-Moliere-DT}), overshoot the theoretical curves, not being reproduced by any of the most commonly used potentials sufficiently well (as was mentioned above, the discrepancies are primarily in $\chi_0$). 
Furthermore, the prediction by Moli\`{e}re potential appears to be closer to those data than HF, although in other cases it was vice versa. 
In this regard, one can consider two basic possibilities. 

The first one is to recall that the experimental procedure for determination of $\overline{\chi}$ consists in fitting the measured angular distribution of transmitted particles by a superposition of Gaussian-related functions, each of which is associated with a certain particle fraction (channeled, dechanneled or  VR). But for negatively charged VR particles, the angular distribution has a rather long `tail' towards the crystal bending due to ``orbiting'' of particles at the tops of broad interplanar potential maxima \cite{Bond-VR-PRA}. Such particles can be misidentified as not belonging to the VR fraction (instead being included to the quickly dechanneled and multiply scattered fraction), and their subtraction increases the mean VR angle.\footnote{It may be mentioned yet that experiment \cite{Wistisen-Wienands} was performed with a rather thin crystal ($L=60\,\mu$m), whereas herein we presume validity of the thick-crystal limit. Nonetheless, condition (\ref{L>>Deltaz}) for it is fulfilled. Besides that, boundary effects are not expected to be strictly positive, and thus cannot explain the stable excess of the data over the theory predictions.} 

Secondly, it should be borne in mind that even though the noninteracting-atom approximation is rather good for third-row chemical elements (much better than for second-row elements \cite{Nishibori,Alatas}), covalent binding effects can nonetheless somewhat alter the interplanar potential \cite{Endo,Lu-Zunger,Nishibori}, and hence  have some impact on $\chi_0$. 
Recalling also (see Fig. \ref{fig:g}) that the observed difference of the continuous planar Moli\`{e}re potential from planar HF potential arose due to a difference between their atomic screening functions at large $r$ ($\sim 1$ \AA), where the applicability of the noninteracting-atom approximation generally breaks down, it is not excluded that real deformations of the valence electron density may give rise to observable differences in continuous potentials, as well. 

If one considers a possibility that the good agreement for $\chi_0$ for positively charged particles could be partly fortuitous, and looks for a modification of the interplanar potential such that the value of $\chi_0$ is unaffected for positively charged particles, but increases for negatively charged ones, qualitatively, it may be instructive to refer to the double-parabolic potential model of Sec. \ref{subsubsec:110poschar} [for simplicity, for the case of orientation (110)]. 
There, $\chi_0$ for positively charged particles was expressed by an explicit function $\chi_0\propto \sqrt{V_0(1-\delta)}$ of the potential depth $V_0$ and shape parameter $\delta$ [see Eq. (\ref{meanVRangle-pos110-sqrtV2})]. 
If the latter is increased so that product $V_0(1-\delta)$ remains constant, for positively charged particles $\chi_0$ will not change, whereas for negatively charged ones, it will increase by factor $1+\delta$  [see Eq. (\ref{chi0=thetac(1+delta2)})]. 
Physically, the increase of $\delta$ corresponds to an increase of the potential curvature midway the planes, which from the viewpoint of the Poisson equation does not contradict to formation of covalent bonds in that region. But to achieve quantitative agreement with the data, one needs sizeable $\delta\sim 0.1$, whereas the fitted value (\ref{fit:V0delta}) is significantly lower.
Proper investigation of this issue is beyond the scope of the present paper.

\section{Summary and outlook}

The developed approach for evaluation of the mean volume reflection angle solves the problem of summation over the crystal periods exactly. 
Apart from the apparent convenience for making numerical predictions, it also permits a number of qualitative inferences. 
In particular, it reveals a duality between VR and channeled particles, in the sense  that only under-barrier regions of the effective potential contribute to the mean VR angle. 
The limiting (at $4R_c/R\ll 1$) value of the latter proves to be related with a net critical channeling angle for a straight crystal (its local generalization averaged over the interplanar interval).

%The representation of the mean VR angle in form of a single integral allows to easily determine which regions of the interplanar interval bring dominant contribution. 
%It appears to be particularly sensitive to the continuous potential shape around the tops of the barriers, because they determine the value of $\max V_{\text{eff}}$, and thereby the entire integrand in Eq. (\ref{general-mean-thetaVR}). Therefore, VR of positively charged particles is more sensitive to the potential near the atomic planes, whereas VR of positively charged particles, to the potential midway the planes.  
%It is also noteworthy that due to the square root dependence of the integrand on the interplanar potential, its sensitivity to subtle differences in the potential is enhanced.
%. The sensitivity to the crystal potential leads in turn to sensitivity to the crystal temperature

%Its predictions
% expresses the VR angle in a simple (especially for the mean VR angle) and generic form. Its predictions %are as general as those of Monte-Carlo simulations, but can be obtained immediately, and are not obscured by statistical errors and the dependence on the procedure of the mean VR angle extraction, such as a Gaussian fit. 

It is also noteworthy that in practice, for description of $E$- and $R$-dependencies of $\overline{\chi}$, it may be sufficient to use NLO expansion in the small ratio $R_c/R$ (Sec. \ref{sec:expansion-RcR}). 
That demands the knowledge of only a few empirical constants, calculable numerically based on a realistic interplanar potential for a straight crystal. NLO formulas for orientation (110) and for parabolic interplanar potential, were formerly implemented in the Monte-Carlo code FLUKA \cite{Schoofs}; they can be improved based on the present development. 
%With the account of the appropriate corrections, the analytic approach can compete with  at typical experimental conditions, especially given that the interplanar potentials are not exactly known, and some analytic parameterizations are employed for them, anyway.

Cases of medium $R/4R_c$ and $R\to R_c$ were also studied. 
At $R\lesssim 4R_c$, some guidance can be provided by parabolic potential models of Sec. \ref{sec:models}, offering simple closed-form expressions for $\overline{\chi}$. 
The most accurate predictions, though, are obtained with HF potentials. 

The comparison between the theory and results of experiment \cite{Wistisen-Wienands} on VR of negatively charged particles and crystal orientation (111) displays yet some discrepancy (Sec. \ref{subsec:experim}), the interpretation of which requires further investigation. More data on VR of negatively charged particles, for all crystal orientations are desirable.

%This opens prospects for using VR not only for applied purposes, but also for fundamental studies, on a par with other processes, such as channeling and coherent bremsstrahlung. 
%The optimal interval for that is $5<R/R_c<20$, where the NLO approximation works satisfactorily, and at the same time multiple scattering within the VR region keeps minor \cite{Bond-PRST}.

%Certain issues in this theory still remain in what concerns the accuracy of our knowledge of the interplanar potential. 

%There may also be some sensitivity to the $V_1/V_2$ ratio for orientation (111). 

It is worth recapitulating that in the present paper, VR was everywhere treated as a purely elastic deflection process. In reality, it is accompanied by weakly inelastic and quasi-elastic scattering processes, which need to be understood, as well. In particular, %for negatively charged particles 
they give rise to a non-negligible volume capture probability, due to which the VR efficiency even at high energy can amount to less than $90\%$. 

Another kind of inelastic processes is emission of electromagnetic radiation, which can be intense for ultra-relativistic electrons and positrons passing through crystals \cite{rad,Hasan,Bandiera-VR-rad}.
For radiation at VR, the deviation of the spectrum from the kinematically formed coherent bremsstrahlung in a bent crystal (generated away from the intrinsic VR region) is concentrated in the low-$\omega$ region \cite{Bondarenco-CBBC}. The infrared limit $\omega\to0$ itself is rather trivial, corresponding to factorization of the radiation and scattering probabilities \cite{Bondarenco-CBBC}. Less trivial is the slope of the radiation spectrum at the origin. For the latter quantity, there was derived a generic  expression depending on the detail of the particle trajectory inside the target \cite{Bondarenco-softrad-NLO}. To describe that behavior, however, one needs to go beyond the present theory, which deals only with the final deflection angle.

%\newpage

\appendix

\section{Derivation of formula (\ref{general-mean-thetaVR-max})}\label{app:sqrt-maxV}

In this appendix we derive formula (\ref{general-mean-thetaVR-max}) central for the present paper. 
If one is interested in the mean VR angle alone, rather than in the detail of the angular distribution, it can be evaluated by
interchanging the order of integrations in the double integral arising when Eq. (\ref{thetaVR=lim}) is inserted into Eq. (\ref{const+DeltaEperp}). 
The dependence of the integrand on $E_{\perp}$ is very simple and should thus be manageable exactly. 

\begin{figure}
\includegraphics{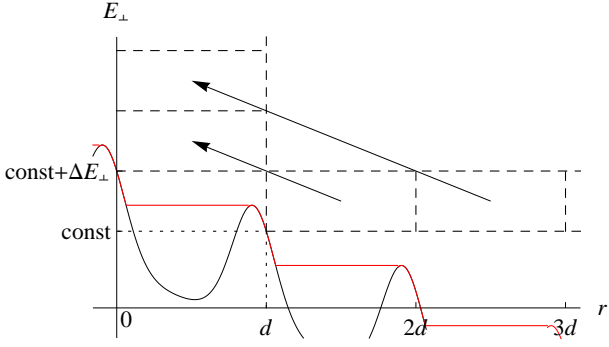}
\caption{\label{fig:region-row} 
Red solid curve, lower boundary of the energetically allowed domain for VR in the $\{r,E_{\perp}\}$ plane. 
Semi-infinite row restricted by the red curve and the horizontal dashed lines -- initial integration region for the mean VR angle. 
Semi-infinite column restricted by the red curve and the vertical dashed lines -- the transformed integration region. 
Arrows show translations of double-full-period cells, not changing the integrand. 
}
\end{figure}

To interchange the order of integrations, one must first take into account that the lower limit $r_{\min}(E_{\perp})$ of $r$ integrations is $E_{\perp}$ dependent. 
Since $r_{\min}(E_{\perp})$ by definition is a maximal $r$ such that $V_{\text{eff}}(r)=E_{\perp}$,  condition $r>r_{\min}(E_{\perp})$ equivalently expresses as an $r$-dependent lower bound for $E_{\perp}$:
\begin{equation}\label{Eperp>minEperp}
E_{\perp}>\min E_{\perp}(r)=\underset{r'\geq r}\max V_{\text{eff}}(r').
\end{equation}
The absolute minimum of $r$ is subsequently inferred by noting that the dependence (\ref{Eperp>minEperp}) is monotonic, whence the minimal $r$ corresponds to the maximal $E_{\perp}$, i.e., to $\text{const}+\Delta E_{\perp}$. 
Without the loss of generality, the origin of coordinate $r$ can be chosen so that
\[
r_{\min}\left(\text{const}+\Delta E_{\perp}\right)=0.
\]
Therewith,
\begin{eqnarray}\label{chi-lim-double-int}
\sqrt{\frac{E}{2}}\overline{\chi}&=&\lim_{r_0\to\infty} \Bigg\{2\sqrt{\frac{Er_0}{R}}\nonumber\\
&\,&-\frac{1}{d}\int_0^{r_0} dr\int_{\underset{r'\geq r}\max V_{\text{eff}}(r')}^{\text{const}+\Delta E_{\perp}}\frac{dE_{\perp}}{\sqrt{E_{\perp}-V_{\text{eff}}(r)}}\Bigg\}.\qquad
\end{eqnarray}
The corresponding integration domain in the $\{r,E_{\perp}\}$ plane is shown by a horizontal band in Fig. \ref{fig:region-row}.

We are now in a position to make use of the periodicity of $V(r)$. 
The integrand of the double integral in (\ref{chi-lim-double-int}) is periodic  both in $r$ and $E_{\perp}$ at constant $E_{\perp}+\frac{E}{R}r$.
If we split the integral with respect to $r$  into a sum of integrals over full periods:
\[
\int_0^{r_0} dr=\int_0^{d} dr + \int_d^{2d} dr+...+ \int_{\lfloor r_0/d\rfloor}^{r_0} dr,
\]
%\begin{eqnarray*}
%\sqrt{\frac{E}{2}}\overline{\chi}&=&\lim_{r'\to\infty}
%\Bigg\{2\sqrt{\frac{Er}{R}}\nonumber\\
%&\,&-\frac{1}{d}\left(\int_0^{d} dr + \int_d^{2d} dr+...+ \int_{...}^{r'} dr\right)\nonumber\\
%&\,&\times\int_{\max V_{\text{eff}}}^{\max V_{\text{eff}}+\Delta E_{\perp}}\frac{dE_{\perp}}{\sqrt{E_{\perp}-V_{\text{eff}}(r)}}\Bigg\}.
%\end{eqnarray*}
each of the partial integrals can be presented as an integral over the same period $[0,d]$, in whose integrand there is yet a compensating shift in the linear component of $V_{\text{eff}}(r)$. 
But those shifts can as well be re-expressed as shifts in $E_{\perp}$, whereby we obtain a sum of integrals over $E_{\perp}$:
\begin{eqnarray*}
\sqrt{\frac{E}{2}}\overline{\chi}=\lim_{r_0\to\infty}\Bigg\{2\sqrt{\frac{Er_0}{R}}\qquad\qquad\qquad\qquad\qquad\qquad\qquad\nonumber\\
-\frac{1}{d}\int_0^{d} dr\sum_{k=0}^{r_0/d} \int_{\underset{r'\geq r}\max V_{\text{eff}}(r')}^{\text{const}+\Delta E_{\perp}}\frac{dE_{\perp}}{\sqrt{E_{\perp}+k\Delta E_{\perp}-V_{\text{eff}}(r)}}\Bigg\}.
\end{eqnarray*}
One observes that they seamlessly combine into a single integral with respect to $E_{\perp}$, with the upper limit $\Delta E_{\perp} r_0/d=Er_0/R$:
\begin{eqnarray}
\sqrt{\frac{E}{2}}\overline{\chi}&=&\lim_{r_0\to\infty}\Bigg\{2\sqrt{\frac{Er_0}{R}}\nonumber\\
&\,&- \frac{1}{d}\int_0^{d} dr\int_{\underset{r'\geq r}\max V_{\text{eff}}(r')}^{\Delta E_{\perp} r_0/d}\frac{dE_{\perp}}{\sqrt{E_{\perp}-V_{\text{eff}}(r)}}\Bigg\}.\qquad
\end{eqnarray}
Graphically this transformation is illustrated in Fig. \ref{fig:region-row}. 
The $E_{\perp}$-integration then performs exactly in a trivial manner, and the limit $r_0\to\infty$ is evaluated to give Eq. (\ref{general-mean-thetaVR-max}).

%\newpage

%.

%\newpage

\section{Derivation of formula (\ref{NLOthetaVR-r*(ER)})}\label{app:NLO-log}

In this appendix we complete the derivation of the logarithmic modification of the generic NLO formula, which was reduced to two integrals in Eq. (\ref{split-r0}), involving an auxiliary ``intermediate-scale'' parameter $r_i$. 
It is expedient first to isolate there the LO contribution (\ref{mean-thetaVR-infty}), complementing the integral $\int_{r_i}^{d}dr \sqrt{\max V-V(r)}$ to an integral over the full period:
\begin{eqnarray*}
\int_{r_i}^{d}dr \sqrt{\max V-V(r)}\simeq \int_{0}^{d}dr \sqrt{\max V-V(r)} \nonumber\\
-\sqrt{\frac{|V''(0)|}{2}}\frac{r_i^2}{2}.
\end{eqnarray*}
Its last term combines with the first term in the right-hand side of Eq. (\ref{split-r0}) to give
\begin{eqnarray*}
\mathfrak{Re}\int_0^{r_i}dr\sqrt{ r^2+\frac{2Ed}{RV''(0)}}-\frac{r_i^2}{2}\qquad\qquad\qquad\qquad\nonumber\\
%=\frac{r_i}{2}\sqrt{ r_i^2+\frac{2Ed}{RV''(0)}}\qquad\qquad\qquad\qquad\qquad\qquad\nonumber\\
%+\frac{Ed}{RV''(0)}\ln\frac{r_i+\sqrt{ r_i^2+\frac{2Ed}{RV''(0)}}}{\sqrt{-\frac{2Ed}{RV''(0)}}}-\frac{r_i^2}{2}\qquad\qquad\nonumber\\
\underset{r_i^2\gg \frac{2Ed}{R|V''(0)|}}\simeq \frac{Ed}{2RV''(0)}\left[1
+\ln\frac{2R|V''(0)|}{Ed}+2\ln r_i\right]
\end{eqnarray*}
(terms quadratic in the auxiliary parameter $r_i$ have cancelled). 
Therewith, Eq. (\ref{split-r0}) becomes
\begin{eqnarray}\label{A1}
\overline{\chi}
=\chi_0
-\frac{1}{R}\sqrt{\frac{E}{|V''(0)|}}\Bigg[1
+\ln\frac{2R|V''(0)|}{Ed}\qquad\qquad\nonumber\\
+2\ln r_i-\frac{ \sqrt{2|V''(0)|}}{d}
\int_{r_i}^{d}dr\frac{r-d}{\sqrt{\max V-V(r)}}
\Bigg]\nonumber\\
+\mathcal{O}\left(\frac{E^2}{R^2}\right).\qquad
\end{eqnarray}

Cancellation of $r_i$ may be accomplished in two ways. One option is to  combine the last two terms in the brackets, using the symmetry of the potential wrt the interplanar interval midpoint $r=d/2$, i.e., $V(d-r)=V(r)$:
\begin{eqnarray*}
\frac{1}{d}\int_{r_i}^{d}dr\frac{ r-d}{\sqrt{\max V-V(r)}}
\equiv-\int_{r_i}^{d/2}\frac{dr}{\sqrt{\max V-V(r)}}\qquad\nonumber\\
=\int_{r_i}^{d/2}d\ln \frac{d}{2r}\frac{r}{\sqrt{\max V-V(r)}}\qquad\qquad\qquad\qquad\qquad\quad\nonumber\\
%=\ln \frac{2r_i}{d}\frac{r_i}{\sqrt{-V(r_i)}}-\int_{r_i}^{d/2}dr\ln \frac{d}{2r}\frac{d}{dr}\frac{r}%{\sqrt{-V(r)}}\nonumber\\
=\frac{1}{\sqrt{|V''(0)|/2}}\ln \frac{2r_i}{d}
-\int_{r_i}^{d/2}\!\!dr\ln \frac{d}{2r}\frac{d}{dr}\frac{r}{\sqrt{\max V-V(r)}},
\end{eqnarray*}
where the latter integral already converges at $r_*\to0$. That leads to the result
\begin{eqnarray}\label{NLO-log-general}
\overline{\chi}
=\chi_0
-\frac{1}{R}\sqrt{\frac{E}{|V''(0)|}}\left(1+\ln\frac{R|V''(0)|d}{2E}\right)\qquad\qquad\quad\nonumber\\
-\frac{\sqrt{2E}}{R}
\int_0^{d/2}dr\ln\frac{d}{2r}\frac{d}{dr}\frac{r}{\sqrt{\max V-V(r)}}
+\mathcal{O}\left(\frac{d^2}{R^2\theta_c^3}\right).\nonumber\\
\end{eqnarray}
%where
%\begin{equation}\label{F0-def}
%    F_0=,
%\end{equation}
%\begin{equation}\label{kappa-def}
%    \kappa=.
%\end{equation}
%As a check, if $|V''(0)|\to\infty$, which corresponds to $V(r)$ having a break in the origin, all the NLO terms except $\kappa$ vanish, and integration by parts in the remaining NLO integral leads back to Eq. (\ref{mean-thetaVR-NLO-sharpedge}).

Alternatively, one can choose such a $r_i=r_*$  that in (\ref{A1}) all the terms in the brackets except the integral cancel:
\[
1
+\ln\frac{2R|V''(0)|}{Ed}+2\ln r_*=0.
\]
That leads to expression (\ref{r*-def}) for $r_*$ and to representation (\ref{NLOthetaVR-r*(ER)}) for $\overline{\chi}$.

\section{Derivation of formula (\ref{meanVRangle-pos110-sqrtV2})}\label{app:B}

Here we derive an approximation for the mean VR angle in a single-well potential described by two joint parabolas, Eq. (\ref{V-piecewise-parab}), suitable under conditions (\ref{R<}) and $d_1\ll d_2$. 
Our starting point is Eq. (\ref{general-mean-thetaVR}).

When the centrifugal potential $-\frac{Er}{R}$ is added to the crystal potential (\ref{mathcalV3-def}), under condition $R\gg R_c$, the location of the maximum of the tilted effective potential shifts from $d$ only slightly: $r_m=d-\frac{Ed_1^2}{8V_1 R}=d-\frac{d_1}{2}\frac{R_c}{R}$. The effective potential value in this maximum equals $V_{\text{eff}}(r_m)=\mathcal{V}_3(r_m)-\frac{Er_m}{R}$. 
Exact evaluation of the integral (\ref{general-mean-thetaVR}) for potential (\ref{V-piecewise-parab}) then gives
\begin{eqnarray}\label{}
\sqrt{\frac{E}{2}}\frac{d}{2}\overline{\chi}\qquad\qquad\qquad\qquad\qquad\qquad\qquad\qquad\quad\nonumber\\
=\int_{d_1/2}^{d-d_1/2}dr\mathfrak{Re}\sqrt{\mathcal{V}_3(r_m)-\mathcal{V}_2(r)+\frac{E(r-r_m)}{R}}\quad\nonumber\\
+\int_{d-d_1/2}^{r_m}dr\sqrt{\mathcal{V}_3(r_m)-\mathcal{V}_3(r)+\frac{E(r-r_m)}{R}}\qquad\nonumber\\
=\mathfrak{Re}\int_{d_1/2}^{d-d_1/2}dr
\Bigg[V_1\left(1-\frac{Ed_1}{4RV_1}\right)^2\qquad\qquad\qquad\nonumber\\
+V_2\left(1-\frac{Ed_2}{4RV_2}\right)^2-V_2\left(\frac{d-2r}{d_2}+\frac{Ed_2}{4RV_2}\right)^2
\Bigg]^{1/2}\nonumber\\
+\int_{d-d_1/2}^{r_m}dr\sqrt{V_1}2\frac{r_m-r}{d_1},\qquad
\end{eqnarray}
%where $a=-\frac{4V_2}{d_2^2}$, $b=\frac{E}{R}+\frac{4V_2d}{d_2^2}$, $c=V_2\left(1-\frac{d^2}{d_2^2}\right)+V_1+V_1\left(\frac{Ed_1}{4V_1R}\right)^2-\frac{Ed}{R}$, 
%$b^2-4ac=\left(\frac{E}{R}-\frac{4V_2}{d_2}\right)^2+\frac{V_1V_2}{d_2^2}\left(4-\frac{Ed_1}{RV_1}\right)^2$
or, evaluating the latter simple integrals,
\begin{eqnarray}\label{B2}
\sqrt{\frac{E}{2}}\frac{d}{2}\overline{\chi}
%=\left(\frac{r}{2}+\frac{b}{4a}\right)\sqrt{ar^2+br+c}\Big|_{r=d-d_1/2}\nonumber\\
%+\frac{b^2-4ac}{8a\sqrt{|a|}}\left(\arcsin\frac{2ar+b^2}{\sqrt{b^2-4ac}}\Big|_{r=d-d_1/2}-\frac{\pi}{2}\right)\nonumber\\
%+\frac{d_1\sqrt{V_1}}{4}\left(1-\frac{Ed_1}{4 RV_1}\right)^2\nonumber\\
&=&\frac{d_1\sqrt{V_1}}{4}\left(1-\frac{Ed_1}{4 RV_1}\right)^2
\nonumber\\
&+&\frac{d_2\sqrt{V_1}}{4}\left(1-\frac{E d_2}{4RV_2}\right)\left(1-\frac{Ed_1}{4RV_1}\right)\nonumber\\
&+&d_2\frac{V_1 \left(1-\frac{Ed_1}{4RV_1}\right)^2+V_2\left(1-\frac{Ed_2}{4RV_2}\right)^2}{4\sqrt{V_2}}\nonumber\\
&\,&\times\left[\pi-\arctan\left(\frac{1-\frac{Ed_1}{4RV_1}}{1-\frac{Ed_2}{4RV_2}}\sqrt{\frac{V_1}{V_2}}\right)\right].\qquad
\end{eqnarray}
Invoking condition (\ref{junction-relation}), wherewith $\frac{Ed_1}{4V_1}=\frac{Ed_2}{4V_2}=R_c$, expression (\ref{B2}) simplifies to
\begin{equation}\label{pos110-hit2parab}
\overline{\chi}
=\sqrt{\frac{V_2}{2E}}
\left(\pi+\sqrt{\frac{V_1}{V_2}}-\arctan\sqrt{\frac{V_1}{V_2}}\right)
\left(1-\frac{R_c}{R}\right)^2.
\end{equation}
Notably, all the $E/R$ dependence here is given by the same factor $\left(1-\frac{R_c}{R}\right)^2$ as for the purely parabolic well, cf. Eq. (\ref{mean-thetaVR-harm-positive-110}). 

In the limit $R/R_c\to\infty$, however, expression (\ref{pos110-hit2parab}) does not exactly tend to $\chi_0$, because it was derived presuming condition (\ref{R<}), where the right-hand side is large but finite.
Its precise limiting value can be obtained by returning to formula (\ref{mean-thetaVR-infty}):
\begin{eqnarray}\label{thetaVRinfty-joint-parab-110}
\chi_0
&=&
\sqrt{\frac{2}{E}}\frac{4}{d}\Bigg[\int_{0}^{d_1/2}dr\sqrt{V_1+V_2-\mathcal{V}_1\left(r\right)}\nonumber\\
&\,&\qquad\quad+\int_{d_1/2}^{d/2}dr\sqrt{V_1+V_2-\mathcal{V}_2\left(r\right)}\Bigg]\nonumber\\
&=&\sqrt{\frac{2V_1}{E}}+\frac{d_2}{d}\sqrt{\frac{2}{EV_2}}\left(V_1+V_2\right)\text{arccot}\sqrt{\frac{V_1}{V_2}},\qquad
\end{eqnarray}
or, with the use of condition (\ref{junction-relation}),
\begin{equation}\label{thetaVRinfty-sqrtV2}
\chi_0=\sqrt{\frac{2V_2}{E}}\left(\frac{\pi}2+\sqrt{\frac{V_1}{V_2}}-\arctan\sqrt{\frac{V_1}{V_2}}\right).
\end{equation}
But as long as $V_1/V_2\lesssim 0.1$, the difference $\sqrt{\frac{V_1}{V_2}}-\arctan\sqrt{\frac{V_1}{V_2}}\approx\frac13\left(\frac{V_1}{V_2}\right)^{3/2}\lesssim 0.01$ with an acceptable accuracy may be neglected both in (\ref{pos110-hit2parab}) and (\ref{thetaVRinfty-sqrtV2}), leaving 
\begin{equation}\label{meanVRangle-pos110-sqrtV2-app}
\overline{\chi}
\approx\frac{\pi}{2}\sqrt{\frac{2V_2}{E}}
\left(1-\frac{R_c}{R}\right)^2.
\end{equation}
That approximation can already be used for arbitrarily large $R$. 

\section*{Acknowledgements}

I thank V. A. Maisheev, A. Mazzolari and U. Wienands for useful discussions. 
This work was supported in part by 
the National Academy of Sciences of Ukraine
(projects 0118U100327 and 0118U006496) and
the Ministry of Education and Science of Ukraine (project 0118U002031).

%\newpage

%\newpage

\end{document}